\documentclass[english,preprintnumbers,amsmath,amssymb,nofootinbib]{revtex4}
\usepackage[T1]{fontenc}
\usepackage[latin9]{inputenc}
\usepackage{color}
\usepackage{float}
\usepackage{amsbsy}
\usepackage{amstext}
\usepackage{graphicx}
\usepackage{esint}

\usepackage{amsmath}

\makeatletter

\@ifundefined{textcolor}{}
{%
 \definecolor{BLACK}{gray}{0}
 \definecolor{WHITE}{gray}{1}
 \definecolor{RED}{rgb}{1,0,0}
 \definecolor{GREEN}{rgb}{0,1,0}
 \definecolor{BLUE}{rgb}{0,0,1}
 \definecolor{CYAN}{cmyk}{1,0,0,0}
 \definecolor{MAGENTA}{cmyk}{0,1,0,0}
 \definecolor{YELLOW}{cmyk}{0,0,1,0}
 }

\usepackage{dcolumn}
\usepackage{bm}
\usepackage{array}

\def\ket{\rangle}
\def\bra{\langle}

\def\epsd{\epsilon_\textrm{d}}

\def\dA{d_\textrm{A}}
\def\dB{d_\textrm{B}}

\def\VB{\bar{V}_\textrm{B}}
\def\eB{\bar{E}_\textrm{B}}
\def\gamB{\bar{\Gamma}}
\def\lamB{\bar{\lambda}_\textrm{B}}

\def\epsEP{\bar{\epsilon}_\textrm{A}}
\def\eEP{\bar{E}_\textrm{A}}
\def\lamEP{\bar{\lambda}_\textrm{A}}
\def\delEP{\Delta_\textrm{EP}}

\def\eps{\epsilon}
\def\lam{\lambda}
\def\del{\delta}
\def\II{I}

\newcommand{\PT}{\mathcal{PT}}

\newcommand{\beq}{\begin{equation}}
\newcommand{\eeq}{\end{equation}}
\newcommand{\beqa}{\begin{eqnarray}}
\newcommand{\eeqa}{\end{eqnarray}}

\makeatother

\usepackage{babel}

\begin{document}

\title{Characteristic dynamics near two coalescing eigenvalues incorporating continuum threshold effects}

\author{Savannah Garmon}
\affiliation{Department of Physical Science,
Osaka Prefecture University,
Gakuen-cho 1-1, Sakai 599-8531, Japan}

\author{Gonzalo Ordonez}
\affiliation{Department of Physics and Astronomy,
Butler University Gallahue Hall,
4600 Sunset Ave.,
Indianapolis, Indiana 46208, USA}

\begin{abstract}
It has been reported in the literature that the survival probability $P(t)$ near an exceptional point where two eigenstates coalesce should generally exhibit an evolution $P(t) \sim t^2 e^{-\Gamma t}$, in which 
$\Gamma$ is the decay rate of the coalesced eigenstate; this has been verified in a microwave billiard experiment [B. Dietz, {\it et al}., Phys. Rev. E {\bf 75}, 027201 (2007)].
However, the heuristic effective Hamiltonian that is usually employed to obtain this result ignores the possible influence of the continuum threshold on the dynamics.  By contrast, in this work we employ an analytical approach starting from the microscopic Hamiltonian representing two simple models in order to show that the continuum threshold has a strong influence on the dynamics near exceptional points in a variety of circumstances.
To report our results, we divide the exceptional points in Hermitian open quantum systems into two cases: at an EP2A two virtual bound states 
coalesce before forming a resonance, anti-resonance pair with complex conjugate eigenvalues, while at an EP2B two resonances coalesce before forming two different resonances.
For the EP2B, which is the case studied in the microwave billiard experiment, we verify the survival probability exhibits the previously reported modified exponential decay on intermediate timescales, but this is replaced with an inverse power law on very long timescales.
    Meanwhile, for the EP2A the influence from the continuum threshold is so strong that the evolution is non-exponential on all timescales and the heuristic approach fails completely.  When the EP2A appears very near the threshold we obtain the novel evolution $P(t) \sim 1 - C_1 \sqrt{t}$ on intermediate timescales, while further away the parabolic decay (Zeno dynamics) on short timescales is enhanced.
\end{abstract}

\received{October 15, 2016}
\revised{April 18, 2017}

\maketitle

\section{Introduction}
\label{sec:intro}

Dissipation in quantum mechanics is a fundamental problem that is relevant to a wide range of dynamical phenomena, including basic problems such as atomic relaxation and nuclear decay.  In many such familiar circumstances, we can generally think of these processes as following a simple exponential decay law.  We note from the outset that the exponential decay is associated with the resonance in quantum mechanics \cite{Siegert39,Nakanishi,SCG78,PPT91,HSNP08,DBP08,Rotter_review,Moiseyev_NHQM,Gadella11,SHO11,HO14,TGKP16,OH,PRSB00}, which can be thought of as a generalized eigenstate that resides outside the ordinary Hilbert space \cite{Nakanishi,SCG78,PPT91,Bohm,Madrid12,MGM05,Moiseyev_NHQM,Hatano_eff,TGKP16,KGTP}.

However, while the exponential process is quite common in nature, it is known that deviations from exponential decay exist in quantum systems at least on very short and extremely long time scales \cite{Fonda,Muga_review}; the short-time deviations typically give rise to parabolic decay $\sim 1 - t^2$ while the long-time deviations give rise to an inverse power law $\sim 1/t^p$, with $p > 0$.  
While the short-time deviation results simply from the exponential form of the time evolution operator in quantum mechanics, the long-time deviations occur as a direct result of the existence of a lower bound on the energy continuum in open quantum systems \cite{Khalfin,Hack}; these continuous degrees of freedom describe the environmental influence or decay channels in these systems.

For many years it had been argued that these deviations might exist but could be difficult to observe; for example, in Ref. \cite{Sudarshan} it is argued that in typical circumstances the long-time deviation does not manifest until after several lifetimes of the exponential decay, by which time the survival probability is so depleted that the process is rendered undetectable.  Despite the challenge, in recent decades both the short time \cite{short_time_expt} and long time deviations \cite{long_time_expt} have been experimentally observed.

However, while the short time and long time deviations are always present, there do exist special circumstances in which the exponential decay effect is modified \cite{EPexpt1d} 
or even vanishes entirely \cite{LR00,Jittoh05,Longhi06,GCV06,GPSS13}.  
In the former case, the usual exponential decay is modified in the vicinity of a so-called \emph{exceptional point} \cite{Kato,Heiss12}, also referred to as a non-Hermitian degeneracy \cite{Berry04}.
Exceptional points (EPs) are discrete non-analytic points in the parameter space of a given Hamiltonian at which two or more eigenvalues coalesce and the usual diagonalization scheme breaks down.  While exceptional points do not appear in the spectrum of closed Hermitian systems, 
they do appear in 
open quantum systems due to the implicit non-Hermitian character of such systems (see \cite{Moiseyev_NHQM,Rotter_review,HO14,TGKP16,Hatano_eff,Berry04,Moiseyev80,HarneyHeiss04,RotterBird15,GRHS12} as well as references within \cite{Moiseyev_NHQM,Rotter_review,RotterBird15}).
Exceptional points have also been studied in recent years in systems that exhibit explicit non-Hermitian characteristics, such as parity-time ($\PT$) symmetric systems \cite{BB98,Znojil01,BBJ02,Bender_review,Jones09} that can be realized experimentally with balanced energy gain and loss that is spatially symmetrized
\cite{RDM05,Kottos_Nature,PTOptExpt2,PTLRC,PTOptExpt3,PT_WGM,Zheng11,GGH15}
or some other effective parity operator combined with time reversal violation
\cite{TUD12,TUD14}.

A given Hamiltonian is no longer diagonalizable at the exceptional point, but instead can only be transformed into a matrix  containing at least one Jordan block \cite{KGTP,Kato,BS96,HeissEP3,GraefeEP3}. The dimension of the Jordan block is given by $N$, the number of coalescing levels. 
Following the convention in the literature, we refer to an EP involving the coalescence of $N$ levels as an EP$N$ \cite{HeissEP3,GraefeEP3,HW16}; however, in the case of two coalescing levels (EP2) considered in this paper we find it useful to introduce two further subcategories, following the convention introduced in Ref. \cite{GGH15}.  In the context of Hermitian open systems, we refer to an exceptional point at which two virtual bound states coalesce before forming a resonance, anti-resonance pair as an EP2A.  Meanwhile, an exceptional point at which two resonance states coalesce before forming two different resonance states is an EP2B  \footnote{Note that this convention was first introduced in Ref. \cite{GGH15} in the context of 
{\it explicitly} non-Hermitian open quantum systems; the primary difference from Ref. \cite{GGH15} and the present work is that exceptional points are not allowed to appear in the first Riemann sheet here in the case of Hermitian open quantum systems.}.  We find it necessary to introduce this distinction in part because, as shown below, the survival probability characteristics in the vicinity of an EP2A and an EP2B are in fact quite different \footnote{We emphasize that even without considering the detailed survival probability dynamics the EP2A and EP2B are clearly distinguished as the former is associated with the emergence of exponential decay (resulting from the appearance of the resonance) \cite{PPT91,GRHS12,OH}, while the latter is associated with the phenomenon of resonance trapping \cite{Rotter_review,RotterBird15,PRSB00,Past07}.}.

Several studies appear in the literature in which the dynamical properties in the vicinity of the exceptional points have been investigated in open systems \cite{EPexpt1d,WKH08,CM11,HeissSS,FMCW14,Hashimoto1} (see also works on the related question of spatial propagation at the EP in certain $\mathcal{PT}$-symmetric systems \cite{PT_EP_dynamics}, as well as enhanced conical diffraction at an EP in a $\mathcal{PT}$-symmetric honeycomb lattice \cite{RKKC12}).
The focus of most of these papers is that in the vicinity of the EP, the usual exponential evolution of the components of the state vector is modified by a linear $t$ term, so that $e^{-iEt} \rightarrow t e^{-iEt}$.  A result of this effect is that the usual exponential decay $e^{- \Gamma t}$ in the survival probability can be modified near the EP2B as $t^2 e^{- \Gamma t}$, which has been verified in a microwave billiard experiment \cite{EPexpt1d} (see also Fig. 13 of Ref. \cite{TUD14}).

Typically this modification of the time evolution near the EP has been demonstrated relying on a heuristic effective Hamiltonian \cite{EPexpt1d,WKH08,CM11}.  While this approach suffices to capture the modified exponential dynamics, at least in the case of the EP2B, it has a significant drawback in that the details of the environmental influence (background continuum) are washed out.  In particular, the influence of the continuum threshold is not considered in this approach.  However, it has been demonstrated in a variety of studies that the influence of the continuum threshold on both exponential \cite{TGKP16,resonator_theory,resonator_expt,John90,QBIC} and non-exponential processes \cite{LR00,Jittoh05,GCV06,GPSS13} can be quite significant in open quantum systems \footnote{We note the influence of the threshold has also recently been studied in certain nuclear scattering reactions; see, for example, [M. Odsuren, Y. Kikuchi, T. Myo, M. Aikawa, and K. Kat\={o}, Phys. Rev. C {\bf 92}, 014322 (2015)] and references within.}.

By contrast, in this paper we approach the time evolution problem at the exceptional point starting from a specific microscopic model Hamiltonian and by only applying methods that incorporate the influence of the continuum.  
For models describing certain open quantum systems, namely the tight-binding model, we rely on the formalism based on the quadratic eigenvalue problem developed in Ref. \cite{HO14}.  In this approach the discrete sector of the system is again eventually described by an effective matrix, but this formalism is obtained by a projection operator technique that treats the continuum exactly \cite{Hatano_eff}.  Relying on this method, we demonstrate that accounting for the influence of the continuum threshold seems particularly crucial in describing the dynamics in the vicinity of the EP2A, which exhibits non-exponential decay on all timescales.  Meanwhile for the dynamics near the EP2B, the pole approximation may suffice on intermediate timescales; however we show that the usual inverse power law decay dominates the dynamics asymptotically.

We evaluate the dynamics for these two cases relying on two representative models.  
The simplest model of an open quantum system that contains an EP2A (at least, as far as we are aware) 
is Model I, described by the Hamiltonian
\beq
  H_\textrm{I}
  	= \epsd d^\dagger d - b \sum_{j=1}^{\infty} \left( c_j^\dagger c_{j+1} + c_{j+1}^\dagger c_j \right)
		- g \left( c_1^\dagger d + d^\dagger c_1 \right)
	.
\label{ham.model.I.intro}
\eeq
Model $\textrm{I}$ consists of a semi-infinite tight-binding chain with an endpoint impurity; here $c_j^\dagger$ is the creation operator at the $j$th site along the chain, for which $b$ gives the strength of the nearest-neighbor coupling, while $d^\dagger$ is the creation operator at the impurity site.  The impurity is coupled to the endpoint of the chain $c_1^\dagger$ with coupling strength $g$ and has an adjustable chemical potential $\epsd$.  However, since the parameter space of Model I does not contain an EP2B, we turn for this case to Model II with the Hamiltonian
\begin{equation}
  H_\textrm{II}
  	= 
	- V \left( \dA^\dagger \dB + \dB^\dagger \dA \right)
	- g \left( c_{1}^\dagger \dB + \dB^\dagger c_{1} \right)
	- b \sum_{j = 1}^\infty \left( c_j^\dagger c_{j+1} + c_{j+1}^\dagger c_j \right)
		,
\label{ham.model.II.intro}
\end{equation}
which consists of a semi-infinite tight-binding chain with a double-impurity (or qubit) coupled at the endpoint of the chain \cite{DBP08}.  Again $c_j^\dagger$ is the creation operator at the $j$th chain site while the qubit operators $\dA^\dagger, \dB^\dagger$ experience the intra-qubit coupling $V$ and the $\dB^\dagger$ site is again coupled to the endpoint of the chain with strength $g$.
Each of these models can be considered as variations of the Friedrichs model that has been used to describe resonance phenomena in open systems \cite{PPT91,Gadella11}.  While our results for these models illustrate likely near-universal characteristic properties for the dynamics near the EPs, our goal is not necessarily to claim the results are universal but more so to emphasize the importance of considering the continuum threshold in the problem, particularly in the case of the EP2A.

In the following section we introduce projection operators for Model I in order to obtain the exact effective Hamiltonian for the discrete spectrum of the model; we further demonstrate the presence of an EP2A in the parameter space of this model.
In Sec. \ref{sec:models.GEP} we map the (nonlinear) discrete effective Hamiltonian to a generalized linear eigenvalue problem that is in one-to-one correspondence with the discrete spectrum, following the method of Ref. \cite{HO14}.  Re-writing the problem in this way enables us to demonstrate in Sec. \ref{sec:models.Jordan} that the eigenvalue problem for the discrete sector can be reduced to Jordan block form at the EP2A; this formalism further enables us to evaluate the dynamics near the EP2A in Sec. \ref{sec:surv}.  
Here we find that the simple pole approximation does not provide a useful description for the evolution near the EP2A on any time scale or for any parameter values.  Instead, the dynamics are strongly determined by the proximity of the exceptional point to the continuum threshold: when the EP2A appears close to the threshold, the survival probability follows a novel $1 - C_1 \sqrt{t} + C_2 t$ evolution on intermediate timescales.  Meanwhile, when the EP2A appears somewhat further from the threshold the short time parabolic decay is strongly enhanced.  In either case, the usual $t^{-3}$ power law decay exerts itself on longer timescales.  

Then in Sec. \ref{sec:II.ham} we briefly outline a similar calculation for the EP2B appearing in Model II.  In this case we find that the pole approximation accurately describes the time evolution on intermediate time scales, but is again replaced by the $t^{-3}$ evolution for long times.  

Finally, in App. \ref{app:models.encircle} we study the topological structure of the parameter space of Model I by demonstrating that the two eigenstates are mapped into one another on parametrically encircling the EP2A.  We briefly discuss this intriguing mathematical property in relation to recent experiments in which dynamical (physical) encirclement of the EP2B has successfully been demonstrated (encirclement of an EP2A has received relatively less attention in the literature).  Some details of the calculations from the main text appear in three further appendices.




\section{Exact effective Hamiltonian and generalized eigenvalue problem for Model I}\label{sec:I}

In the absence of interaction terms, we can write the Hamiltonian Eq. (\ref{ham.model.I.intro}) for Model I in the single particle picture \footnote{Particularly in light of recent experimental work \cite{many_body_expt} demonstrating that one can distinguish between the dynamics of fermion and boson systems even in the absence of interactions \cite{LDV12}, we leave the extension of the present problem to the many body case to future work.} as
\beq
  H_\textrm{I}
  	= \epsd  | d \ket \bra d | - \sum_{j=1}^{\infty} \left( | j \ket \bra j+1 | + | j+1 \ket \bra j | \right)
		- g \left( | 1 \ket \bra d | + | d \ket \bra 1 | \right)
\label{ham.model.I}
\eeq
with $| d \ket = d^\dagger | 0 \ket$ and $| j \ket = c_j^\dagger | 0 \ket$ as the single particle basis kets,
and where we have set the energy units in terms of $b=1$.
Model I consists of a semi-infinite tight-binding chain coupled with an endpoint impurity; the adjustable parameters in this case are the impurity energy $\epsd$ and the endpoint coupling $g$.  
We assume $g < 1$ because, as seen below, this is the case for which two EP2As can be realized.
Model I is perhaps the ideal model for the study of the EP2A as the discrete spectrum consists of only two states, which coalesce at the exceptional points.  
In the time evolution problem studied in Sec. \ref{sec:surv} we will assume the particle is initially trapped in the endpoint impurity $| d \ket$ and observe how the EP2A influences the evolution from this initial condition.

To begin our analysis for Model I, we first aim to obtain a compact effective Hamiltonian describing the spectrum in the discrete sector following the Feshbach projection operator technique \cite{Rotter_review,Hatano_eff,Feshbach_1,Feshbach_2,SR03}.   After we take the continuum limit for the semi-infinite tight-binding chain in Eq. (\ref{ham.model.I}) the discrete sector for Model I consists simply of the impurity state $| d \ket$; hence we introduce the discrete sector projection operator $P$ as
\beq
  P
  	= | d \ket \bra d |
\label{I.P}
\eeq
while the projection operator $Q$ associated with the continuum is given by
\beq
  Q
  	= 1 - P
	=  \sum_{j=1}^\infty | j \ket \bra j |
	.
\label{I.Q}
\eeq
Below we will obtain an energy-dependent effective finite dimensional Hamiltonian $H_\textrm{eff} (E)$ for Model I, which yields the generalized discrete eigenvalues $E_j$ associated with the eigenkets $| \psi_j \ket$ as solutions of the equation
\beq
  H_\textrm{eff} (E_j) P | \psi_j \ket 
  	= E_j \left( P | \psi_j \ket \right)
	.
\label{H.eff.schr}
\eeq
Note that since $H_\textrm{eff} (E)$ itself depends on the energy, the dimension of this operator does not, in general, match the dimension of the solution space, and further the eigenvalue solutions of this equation technically belong to \emph{differing} versions of this Hamiltonian.
Hence, while this equation can correctly predict the coalescence of the eigenvalues at a given EP, it cannot by itself properly describe the coalesce of the associated eigenvectors. 
However, following the method in Ref. \cite{HO14} we can map the above equation to an equivalent generalized eigenvalue problem, which does include all of the discrete eigenvalues as solutions of a single matrix equation.  
As revealed below, this formalism can be used to describe all of the non-analytic properties of the system in the vicinity of the exceptional point, while treating the continuum exactly.

We obtain the effective Hamiltonian for Model I 
based on the Siegert boundary condition, which consists of an outgoing wave from the impurity region written as
\beq
  \psi (x) \equiv
  	\bra x | \psi \ket
	=  \left\{ \begin{array}{ll}
		\psi_\textrm{d}				& \mbox{for $x = \textrm{d}$}   	\\
		C e^{ikx}					& \mbox{for $x \ge 1$}
	\end{array}
	\right.  
	.
\label{I.outgoing}
\eeq
We analyze the Schr\"odinger equation $H | \psi \ket = E | \psi \ket$ for any site $x \ge 2$ as 
$\bra x | H | \psi \ket = E \bra x | \psi \ket$, which gives
\beq
  - \psi \left( x - 1 \right) - \psi \left( x + 1 \right)
  	= E \psi \left( x \right)
	.
\label{sch.eq.chain}
\eeq
Applying our plane wave solution $\psi (x) = C e^{ikx}$ from Eq. (\ref{I.outgoing}) we obtain the continuous eigenvalue
\beq
  E \left( k \right)
  	= - 2 \cos k
		.
\label{cont.disp}
\eeq
The scattering continuum is hence defined in the range $|E(k)| \le 2$ on the domain $k \in [ 0, \pi]$ for the semi-infinite chain.  For a given discrete solution $\psi_j (x)$, Eq. (\ref{cont.disp}) also connects the wave vector $k_j$ with the relevant eigenvalue $E_j$, where both $k_j$ and $E_j$ are, in general, complex.  Evaluating  Eq. (\ref{I.outgoing}) at $x=1$ and solving the Schr\"odinger equation in the impurity region for sites $x = \textrm{d}, 1$,  yields the coupled equations
\beqa
  - \psi (2) - g \psi_\textrm{d} & = & E \psi (1)
  					\label{I.sch.eq.imp.1} \\
  \epsd \psi_\textrm{d} - g \psi (1) & = & E \psi_\textrm{d}
  	.
\label{I.sch.eq.imp.2}
\eeqa

We can use Eq. (\ref{I.outgoing}) to write $\psi(2) = e^{ik} \psi(1)$, which in turn allows us to rewrite Eq. (\ref{I.sch.eq.imp.1}) as $e^{-ik} \psi (1) = g \psi_\textrm{d}$ after applying Eq. (\ref{cont.disp}).  We can plug this expression into Eq. (\ref{I.sch.eq.imp.2}) to finally obtain the effective eigenvalue equation (\ref{H.eff.schr}) in which $H_\textrm{eff} (E)$ here takes the form
\beq
  H_\textrm{eff} (E)
  	= \epsd - g^2 e^{ik}
		,
\label{I.H.eff}
\eeq
where $E$ depends on $k$ according to Eq. (\ref{cont.disp}).
Note that $H_\textrm{eff}$ in general is a matrix with dimension equal to that of the discrete subspace $P$; hence for Model I with $P = | d \ket \bra d |$ the effective Hamiltonian Eq. (\ref{I.H.eff}) is just a scalar operator in this case.

	%
	%

Applying our expression for the effective Hamiltonian Eq. (\ref{I.H.eff}) in the Schr\"odinger equation (\ref{H.eff.schr}) we obtain the dispersion equation for the discrete eigenvalues in quadratic form.  The two eigenvalue solutions are given by
\beq
  E_{\pm}
  	= \frac{\epsd \left( 2 - g^2 \right) \pm g^2 \sqrt{\epsd^2 - 4 \left( 1 - g^2 \right)} }
			{2 \left( 1 - g^2 \right)}
\label{I.E.pm}
\eeq
while the associated wave vectors $k_\pm$ are given by $E_\pm = - 2 \cos k_\pm$. From Eq. (\ref{I.E.pm}) we can immediately find the location of the two exceptional points (EP2As) in parameter space where the two discrete eigenvalues coalesce as 
$\epsd = \pm 2 \sqrt{1 - g^2}$, which are the branch points of the square root appearing in these two equations   (keep in mind we assume $g<1$ so that these exceptional points are real-valued).  
Hence, for $| \epsd | < 2 \sqrt{1 - g^2}$ it is easy to see that $E_\pm$ yield an anti-resonance/resonance pair with complex conjugate eigenvalues, while they become real-valued on the other side of the EP2As.  To make a more precise statement, it can be shown that the eigenvalues evolve in the parameter space as follows:
\beq
  E_\pm
	=  \left\{ \begin{array}{ll}
		\mbox{resonance, anti-resonance}		& \ \ \  \mbox{for $| \epsd | < 2 \sqrt{1 - g^2}$}   	\\
		\mbox{two virtual bound states}		& \ \ \  \mbox{for $2 \sqrt{1 - g^2} < | \epsd | < 2 - g^2$}	\\
		\mbox{one bound, one virtual bound state}	& \ \ \  \mbox{for $2 - g^2 < | \epsd | $}
	\end{array}
	\right.  
	.
\label{I.E.pm.par}
\eeq
The properties of the spectrum throughout the parameter space are studied in detail in Ref. \cite{GRHS12}.

From this point forward, we will primarily focus on the properties of the system very near the 
\emph{lower} exceptional point, which we denote by $\epsd = \epsEP$ with
\beq
  \epsEP
  	\equiv - 2 \sqrt{1 - g^2}
\label{eps.EP}
\eeq
(essentially the same analysis could be applied for the upper EP2A).  Hence, we will primarily study the case $\epsd \gtrsim \epsEP$ for which the spectrum gives a resonance/anti-resonance pair on one side of the EP2A, and the case $\epsd \lesssim \epsEP$ for which it gives two virtual bound states on the other side.
Note that at the lower EP2A, the energy eigenvalues coalesce at
\beq
  E_+ = E_- =
  \eEP
  	\equiv - \frac{2 - g^2}{\sqrt{1 - g^2}}
	.
\label{E.EP}
\eeq



\subsection{Mapping to the generalized eigenvalue problem for Model I}\label{sec:models.GEP}

The previous eigenvalue equation (\ref{H.eff.schr}) with the effective Hamiltonian Eq. (\ref{I.H.eff}) can be mapped onto the quadratic eigenvalue problem \cite{TM01} in terms of $\lambda = e^{i k}$ as
\beq
  \left[ \left( 1 - g^2 \right) \lambda^2 + \epsd \lambda  + 1 \right] P | \psi \ket
  	= 0
	;
\label{I.QEP.eq}
\eeq
from which we immediately obtain the eigenvalues in terms of $\lambda$ as
\beq
  \lambda_\pm
  	= \frac{- \epsd \mp \sqrt{\epsd^2 - 4 \left( 1 - g^2 \right)} }{2 \left( 1 - g^2 \right)}
\label{I.lambda.pm}
\eeq
as well as the useful identity
\beq
  \lambda_+ \lambda_- \left( 1 - g^2 \right)
  	= 1
	.
\label{I.lambda.pm.ident}
\eeq
While of course the solutions presented in Eq. (\ref{I.lambda.pm}) are entirely equivalent to those previously reported in Eq. (\ref{I.E.pm}) and therefore provide no additional information about the spectrum, equation (\ref{I.QEP.eq}) will allow us to rewrite the problem in terms of a generalized eigenvalue equation such that the eigenkets are solutions of a single distinct linear operator, shown below.  
Note that at the exceptional points $\epsd = \pm 2\sqrt{1-g^2}$, the $\lambda$ eigenvalues coalesce at $\lambda_+ = \lambda_- = -2/\epsd$.  For the lower EP2A (focus of the present development) we write this as
\beq
  \lambda_+ = \lambda_- =
  \lamEP
  	\equiv - \frac{2}{\epsEP}
	= \frac{1}{\sqrt{1 - g^2}}
	.
\label{lambda.EP}
\eeq

Noting that Eq. (\ref{I.QEP.eq}) above takes the form of the quadratic eigenvalue problem, we can apply techniques from Refs. \cite{TM01,SM_book} to re-write this as
\beq
  	\left[ \begin{array}{cc}
		-\lambda 	&	1	\\
		1		&	\epsd + \left( 1 - g^2 \right) \lambda
	\end{array} \right]
	\left[ \begin{array}{c}
		P | \psi \ket		\\
		\lambda P | \psi \ket
	\end{array} \right]
		= 0
	.
\label{I.GEP.eq}
\eeq
This equation takes the form of the generalized linear eigenvalue problem
\beq
  \left( F - \lambda G \right) | \Psi \ket 
  	= 0
\label{GEP}
\eeq
with the model-dependent $F$ and $G$ matrices given in this case by
\beq
  F
  	= \left[ \begin{array}{cc}
		0  	&  1		\\
		1  	&  \epsd
		\end{array} \right]
				\ \ \ \ \ \ \ \ \ \ \ \ \ \ \ \ \ \ \ \ 
  G
  	= \left[ \begin{array}{cc}
		1  	&  0		\\
		0  	&  -1 + g^2
		\end{array} \right]
\label{I.A.B}
\eeq
as well as
\beq
  | \Psi \ket
  	\equiv \left[  \begin{array}{c}
		P | \psi \ket	\\
		\lambda P | \psi \ket
		\end{array} \right]
	.
\label{Psi.defn}
\eeq
We can immediately re-obtain the $\lambda$-eigenvalues Eq. (\ref{I.lambda.pm}) from this generalized eigenvalue equation.  However, the real advantage to Eq. (\ref{I.GEP.eq}) is that we have now re-written the eigenvalue equation as a matrix equation in which the dimension is in one-to-one correspondence with the number of discrete eigenvalues.  This will enable us to write the Jordan block structure at the EP2A and further to evaluate the survival probability in the vicinity of the exceptional point without introducing any {\it a priori} approximation.


\subsection{Diagonalization scheme for the generalized eigenvalue problem}\label{sec:models.diag}

We here present the diagonalization scheme for Model I (away from the exceptional points) in the context of the generalized eigenvalue problem Eq. (\ref{GEP}).  The general expression for the survival amplitude that we will use in Sec. \ref{sec:surv.int} can in turn be obtained based on this diagonalization scheme (see Ref. \cite{HO14} for details).

The generalized eigenvalue problem in Eq. (\ref{GEP}) appears in terms of the right-eigenvector 
$| \Psi \ket$.  It is natural to introduce the left-eigenvectors $| \tilde{\Psi} \ket$ satisfying
\beq
  \bra \tilde{\Psi} | \left( F - \lambda G \right)
  	= 0
	.
\label{GEP.trans}
\eeq
If we then normalize the eigenvector set such that
\begin{equation}
  \bra \tilde{\Psi}_i \left| G \right| \Psi_j \ket = \delta_{i,j}
  	,
\label{B.norm}
\end{equation}
we obtain the natural diagonalization for the $F$ matrix as
\begin{equation}
  \bra \tilde{\Psi}_i \left| F \right| \Psi_j \ket = \lambda_j \delta_{i,j}
  	.
\label{A.diag}
\end{equation}
We can immediately obtain the general form of the left-eigenvectors $\bra \tilde{\Psi} |$ by taking the transpose of Eq. (\ref{GEP}) and comparing the result to Eq. (\ref{GEP.trans}).  Noting that 
$F^\textrm{T} = F$ and $G^\textrm{T} = G$ we find that the left-eigenvectors are simply the transpose of the right-eigenvectors $\bra \tilde{\Psi} | = | \Psi \ket^\textrm{T}$.

To find the appropriate normalization for the two generalized Model I eigenstates $| \Psi_\pm \ket$ we
plug the explicit form of $G$ from Eq. (\ref{I.A.B})
into Eq. (\ref{B.norm}) to obtain the condition
\beq
  \left[ 1 - \left( 1 - g^2 \right) \lambda_\pm^2 \right]
  		 \bra d | \psi_\pm \ket^2
	= 1
	,
\label{I.norm.cond}
\eeq
where we have used Eq. (\ref{Psi.defn}) and 
$\bra \tilde{\psi}_\pm | P | \psi_\pm \ket = \bra \tilde{\psi}_\pm | d \ket \bra d | \psi_\pm \ket = \bra d | \psi_\pm \ket^2$.
Taking the positive square root of Eq. (\ref{I.norm.cond}) we find
\beq
  \bra d | \psi_\pm \ket
  	= \frac{1}{\sqrt{1 - \left( 1 - g^2 \right) \lambda_\pm^2}}
	\equiv \beta_\pm
	.
\label{I.norm}
\eeq
From Eq. (\ref{I.lambda.pm}), we notice immediately that the norm $\beta_\pm$ diverges at the EP2A when $\lambda_+ = \lambda_- = (1-g^2)^{-1/2}$; this is a first indication that the usual analytic properties of the system break down at the exceptional point.

To complete the diagonalization of the matrix $F$, we write the right unitary transformation matrix
\beq
  U
  	= \left[ \begin{array}{cc}
		\bra d | \Psi_+ \ket	&  \bra d | \Psi_- \ket
		\end{array} \right]
	= \left[ \begin{array}{cc}
		\beta_+			&  \beta_-		\\
		\lambda_+ \beta_+	&  \lambda_- \beta_-
		\end{array} \right]
	,
\label{I.U}
\eeq
as well as the appropriate left conjugate transformation matrix obtained as $\tilde{U} = U^{-1} G^{-1}$,
which guarantees 
\beq
\tilde{U} G U = I_2
	.
\label{B.ident}
\eeq
We apply the same unitary transformation to diagonalize the $F$ matrix according to
\beq
  \tilde{U} F U
	= \left[ \begin{array}{cc}
		\lambda_+	&  0		\\
		0			& \lambda_-
		\end{array} \right]
	\equiv \Lambda
	.
\label{A.Lambda}
\eeq
As already pointed out, this diagonalization scheme fails at the exceptional points where the eigenstates coalesce and the norm reported in Eq. (\ref{I.norm}) diverges.
Instead, the matrix $F$ can only be reduced to Jordan block form in this case, as shown in the next subsection.


\subsection{Jordan block structure at the exceptional point}\label{sec:models.Jordan}

At the lower exceptional point $\epsd = \epsEP$ in Model I, the $\lambda$ eigenvalues coalesce at 
$\lambda_+ = \lambda_- = \lamEP$ and the $F$ and $G$ matrices take the form
\beq
  F (\epsd = \epsEP)
  	= \left[ \begin{array}{cc}
		0  	&  1		\\
		1  	&  \epsEP
		\end{array} \right]
					,
				\ \ \ \ \ \ \ \ \ \ \ \ \ \ \ \ \ \ \ \ 
  G (\epsd = \epsEP)
  	= \left[ \begin{array}{cc}
		1  	&  0		\\
		0  	&  - \epsEP^2 / 4
		\end{array} \right]
		.
\label{I.A.B.EP}
\eeq
The corresponding eigenstates also coalesce as $| \Psi_\textrm{EP} \ket$ to satisfy a single generalized eigenvalue equation
\beq
  F | \Psi_\textrm{EP} \ket
  	= \lamEP G  | \Psi_\textrm{EP} \ket
	.
\label{Psi.EP.eqn}
\eeq
However, as noted previously, the eigenvector norm Eq. (\ref{I.norm}) diverges at the exceptional point as 
$\beta_\pm \rightarrow \infty$; further, the  coalesced eigenvector exhibits self-orthogonality \cite{Moiseyev_NHQM}.  Renormalizing the coalesced eigenvector in terms of the modified norm $\mu$ we have
\beq
  | \Psi_\textrm{EP} \ket
  	= \mu \left[  \begin{array}{c}
		1	\\
		\lamEP
		\end{array} \right]
	= \mu \left[  \begin{array}{c}
		1	\\
		- \frac{2}{\epsEP}
		\end{array} \right]
	,
\label{I.Psi.EP}
\eeq
from which we immediately see the self-orthogonal character as
\begin{equation}
  \bra \tilde{\Psi}_\textrm{EP} \left| G \right| \Psi_\textrm{EP} \ket = 0
  	.
\label{self.orthogonality}
\end{equation}


Clearly the single eigenvector $| \bar{\Psi}_\textrm{EP} \ket$ does not span the basis of the 2x2 $F$ and $G$ matrices, which is yet another indication that the system is no longer diagonalizable at the exceptional point.  Instead, the system can only be reduced to Jordan block form; as such Eq. (\ref{Psi.EP.eqn}) is complemented by \cite{GraefeEP3,BS96}
\beq
  F | \Phi_\textrm{EP} \ket
  	= \lamEP G  | \Phi_\textrm{EP} \ket + G | \Psi_\textrm{EP} \ket
	,
\label{Phi.EP.eqn}
\eeq
where $| \Phi_\textrm{EP} \ket$ is the so-called pseudo-eigenvector.  Equations (\ref{Psi.EP.eqn}) and (\ref{Phi.EP.eqn}) together constitute the Jordan chain relations.

Writing the pseudo-eigenvector as $| \Phi_\textrm{EP} \ket = \left[ \phi_1, \phi_2 \right]^\textrm{T}$, there are now three unknowns in the problem: $\mu, \phi_1, \phi_2$.  While the Jordan chain vectors (and hence these three unknowns) are not uniquely determined, we can fix the values of $\phi_1$ and $\phi_2$ by imposing the following conditions on the pseudo eigenvector
\beqa
  \bra \tilde{\Phi}_\textrm{EP} \left| G \right| \Phi_\textrm{EP} \ket = 0
  	,		\label{pseudo.cond.1}		\\
  \bra \tilde{\Phi}_\textrm{EP} \left| G \right| \Psi_\textrm{EP} \ket = 1
  	.
\label{pseudo.cond.2}
\eeqa
These conditions follow rather naturally in response to Eq. (\ref{self.orthogonality}) and are similar to those appearing in Ref. \cite{GraefeEP3}.  Applying these constraints we find $\phi_2 = 1 / 2\mu$ and $\phi_2 = 1 / \mu \epsEP$.  We can now write a rotation matrix $R$ as
\beq
  R
  	= \left[ \begin{array}{cc}
		| \Psi_\textrm{EP} \ket	&  | \Phi_\textrm{EP} \ket
		\end{array} \right]
	= \left[ \begin{array}{cc}
		\mu					&  \frac{1}{2 \mu}		\\
		- \frac{2 \mu}{\epsEP}	&  \frac{1}{\mu \epsEP}
		\end{array} \right]
	,
\label{I.R}
\eeq
which essentially takes the place of the usual diagonalizing matrix $U$ at the exceptional point.
Defining as well the left partner rotation matrix $\tilde{R} = R^{-1} G^{-1}$, we can now put the $F$ matrix at the exceptional point into Jordan normal form by writing
\beq
  \tilde{R} F R
  	= \left[ \begin{array}{cc}
		- \frac{2}{\epsEP}		&  \frac{2}{\mu^2 \epsEP}		\\
		0					& - \frac{2}{\epsEP}
		\end{array} \right]
	  = \left[ \begin{array}{cc}
		\lamEP				&  1		\\
		0					& \lamEP
		\end{array} \right]
\eeq
where we have chosen $\mu = \sqrt{2/ \epsEP}$ in the last step.

As a further mathematical property, we show in App. \ref{app:models.encircle} that the eigenstates $| \Psi_+ \ket$ and $| \Psi_- \ket$ are mapped into one another upon topological encirclement of one or the other (but not both) exceptional points, except that one or the other eigenstate will pick up an additional 180$^\circ$ phase change.  This property of EP2s has already been demonstrated over a decade ago 
in a series of quasi-static microwave cavity experiments \cite{EPexpt1a,EPexpt1b,EPexpt1c,TUD11,TUD14} (see also Ref. \cite{Lee09}), although a true dynamical encirclement is a more complex problem \cite{GMM13,Rotter15} that has only been achieved in experiment very recently for the EP2B \cite{HeissNat,RotterNat,XuNat}.  
In App. \ref{app:models.encircle} we demonstrate the topological encirclement for the relatively less-studied EP2A case and briefly comment on what the calculation for the dynamical evolution in this case might entail.

In this subsection we have shown that in the generalized eigenvalue problem based on the formalism of Ref. \cite{HO14} the discrete sector of the present system can only be reduced to Jordan block form at the exceptional point.  
For a more generalized treatment of the problem regarding the Jordan block at exceptional points in open quantum systems,
see Ref. \cite{KGTP}.
We now turn to the time evolution problem near the exceptional point for Model I in the next section.

\section{Model I: Survival probability near the EP2A}\label{sec:surv}

In this section we evaluate the survival probability $P(t)$ for Model I assuming that the endpoint impurity 
$| d \ket$ is initially occupied at time $t = 0$.  The survival probability is written as
\beq
P(t)
	= |A(t)|^2
	,
\eeq
which is the squared modulus of the survival amplitude
\beq
A(t)
	= \bra d | e^{-i H_\textrm{I} t} | d \ket
	.
\label{I.surv.amp}
\eeq


As a starting point, it is straightforward to demonstrate that the system follows parabolic decay on very short timescales, regardless of proximity to the EP2A.  To show this, we first simply expand the time evolution operator in Eq. (\ref{I.surv.amp}) for small $t$ (we will quantify this condition more precisely momentarily) as
\beq
A_\textrm{Z}(t)
	\approx 1 -i t \bra d | H_\textrm{I} | d \ket - \frac{1}{2} t^2 \bra d | H_\textrm{I}^2 | d \ket
\label{I.surv.amp.zeno.H}
\eeq
Evaluating $\bra d | H_\textrm{I} | d \ket  = \epsd$ and $\bra d | H_\textrm{I}^2 | d \ket = \epsd^2 + g^2$ for the Hamiltonian in Eq. (\ref{ham.model.I}) we immediately find
\beq
P_\textrm{Z}(t)
	\approx 1 - g^2 t^2
	,
\label{I.surv.prob.zeno}
\eeq
revealing the non-exponential behavior on short times.  We can more carefully quantify the validity of the expansion in Eq. (\ref{I.surv.amp.zeno.H}) by noting that it should hold for $t$ satisfying
\beq
  \left| i t \bra d | H_\textrm{I} | d \ket \right|
	\lesssim 1
	,
\eeq
yielding the condition $t \lesssim T_\textrm{Z}$ in which the timescale $T_\textrm{Z}$ is given by
\footnote{Note this is consistent with the timescale reported for the short time dynamics in 
T. Petrosky and V. Barsegov,
Phys. Rev. E {\bf 65}, 046102 (2002).}
\beq
  T_\textrm{Z}
  	= \frac{1}{\epsd}
	.
\label{T.Z}
\eeq
We emphasize this effect results directly from the exponential form of the time evolution operator and that we made no reference to specific choices for the system parameters in order to derive it.  Hence, the parabolic decay on short times is a universal effect that will appear at an exceptional point just the same as away from it.  While in most cases this effect is rather difficult to observe since it controls the evolution for such a brief period, we notice that the timescale in Eq. (\ref{T.Z}) hints that for small $\epsd$ the effect might be enhanced.  We demonstrate below Eq. (\ref{I.P.band.edge}) that this is indeed the case.

However, to first evaluate the dynamics near the exceptional point on the intermediate and long timescales, we will find it useful to introduce in Sec. \ref{sec:surv.int} an expression for the survival amplitude based on the generalized eigenvalue problem, as originally obtained in Ref. \cite{HO14}.  Relying on this approach, we will show in Sec. \ref{sec:surv.int2} that the time evolution is non-exponential near the EP2A even on intermediate timescales.  Further we will find that the dynamics are in turn strongly influenced by the proximity of the EP2A to the continuum threshold.  In anticipation of this result, we will first briefly review in Sec. \ref{sec:surv.cont} the influence of an isolated virtual bound state on the non-exponential dynamics when it appears near the threshold \cite{GPSS13}.  The dynamics near the EP2A revealed in Sec.  \ref{sec:surv.int2} can be understood as, in part,  resulting from this effect.

\subsection{Continuum threshold influence on non-exponential decay: a brief review}\label{sec:surv.cont}

While the short time deviations from exponential decay in quantum mechanics can be viewed as resulting simply from the form of the evolution operator (as demonstrated above), it is the existence of a lower or upper bound (threshold) on the energy continuum in open systems that results in non-exponential decay on long time scales \cite{Khalfin,Hack}.  Hence it is rather natural that a discrete eigenvalue appearing in the vicinity of the threshold would result in an enhancement of the non-exponential dynamics, including cases in which the exponential decay vanishes completely \cite{LR00,Jittoh05,GCV06,GPSS13,DBP08,LonghiPRA06}.

In particular, it is argued in Ref. \cite{GPSS13} that a virtual bound state appearing near the continuum threshold can have a strong governing influence on the non-exponential dynamics.  The non-exponential dynamics in this case can be divided into two characteristic time zones, separated by the time scale $\Delta_Q^{-1}$ where $\Delta_Q$ is the energy gap between the virtual bound state energy and the threshold.  Specifically it is suggested in Ref. \cite{GPSS13} that under fairly general circumstances a relatively intermediate time scale (the {\it long time near zone}) should appear during which the survival amplitude falls off as $t^{-1/2}$ and hence the survival probability exhibits a $t^{-1}$ decay, while for the asymptotic {\it long time far zone} the amplitude follows $t^{-3/2}$ and the survival probability decays as $t^{-3}$ 
\footnote{We note that in [A. M. Ishkhanyan, V. P. Krainov, Phys. Lett. A {\bf 379}, 2041 (2015)] the authors claim to find a quantitatively different power law decay than those reported in Ref. \cite{GPSS13}; however, they have inappropriately compared the survival {\it amplitude} in their case, which they state falls off as $t^{-1/2}$, with the survival {\it probability} in the near zone in Ref. \cite{GPSS13}, which falls off as $t^{-1}$.   Noting that only the survival probability is actually measurable in experiment, comparing the physical quantity from both cases would yield the same power law $t^{-1}$.}. The latter result for the far zone dynamics is fairly well known in the literature; for example, see Refs. \cite{GCV06,GPSS13,GCMM95,MDS95,Zueco16,DBP08,HO14}.

In Sec. \ref{sec:surv.int2} below we will find that the near zone effect plays an indirect role in determining the intermediate time scale dynamics when the EP2A is close to the threshold.

\subsection{Survival probability in the generalized eigenvalue problem formalism}\label{sec:surv.int}

\begin{figure}
\hspace*{0.05\textwidth}
 \includegraphics[width=0.4\textwidth]{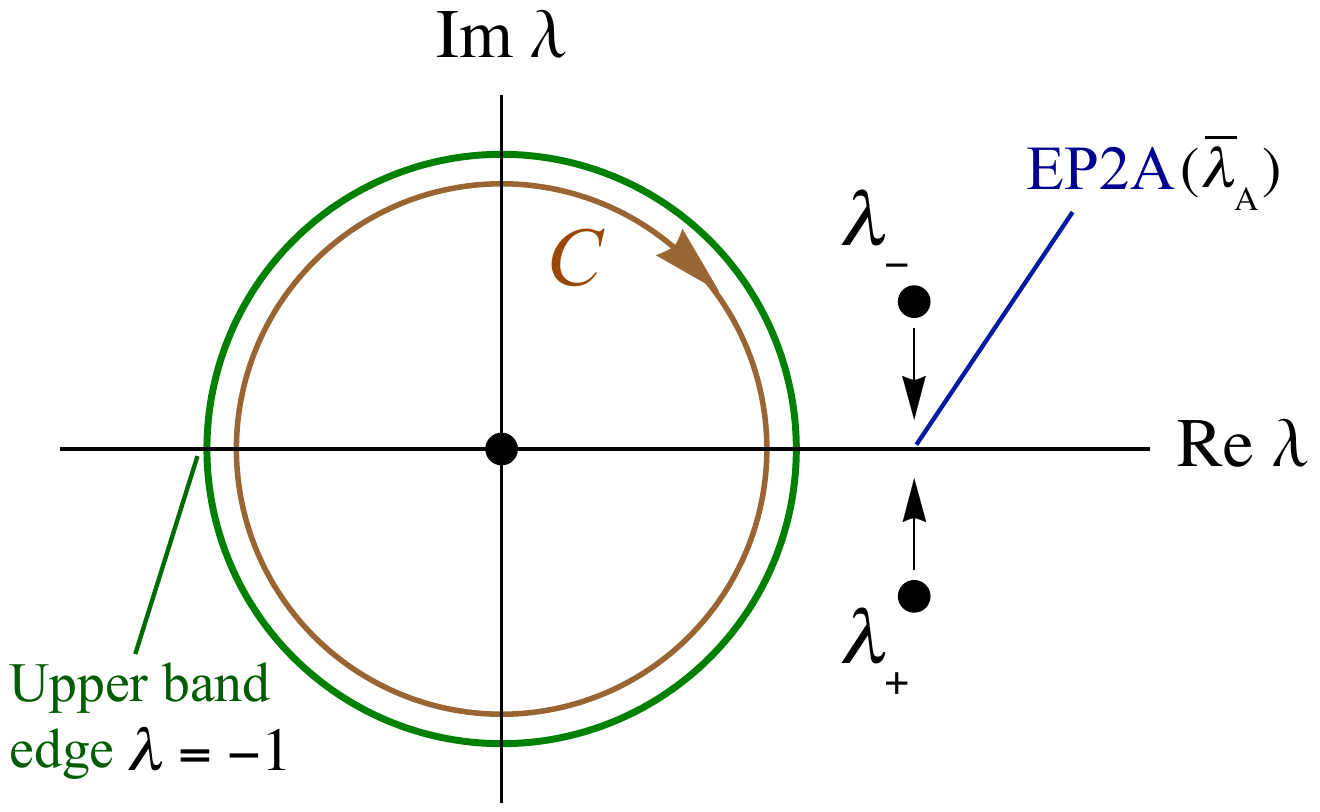}
\hfill
 \includegraphics[width=0.4\textwidth]{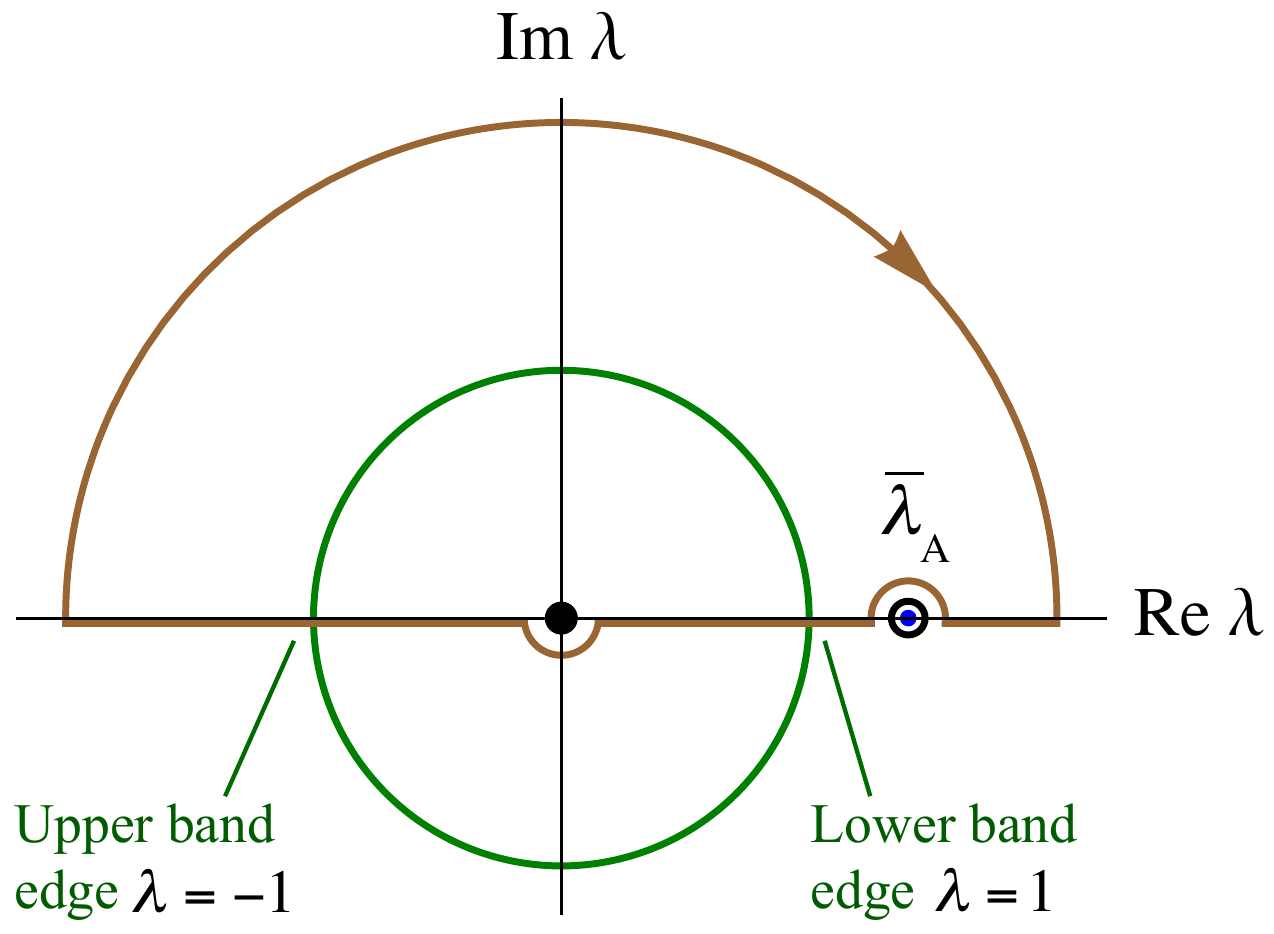}
 \hspace*{0.05\textwidth}
\\
\vspace*{\baselineskip}
\hspace*{0.01\textwidth}(a)\hspace*{0.47\textwidth}(b)\hspace*{0.4\textwidth}
\\
\caption{(color online) Integration contours in the complex $\lambda$ plane.  The continuum coincides with the unit circle, indicated in green; the lower (upper) band edge has been mapped from $E=-2$ ($E=2$) to $\lambda = 1$ ($\lambda = -1$).  An essential singularity appears at the origin, while another occurs at infinity.
(a) The $\lambda_\pm$ eigenvalues approach the coalescent eigenvalue $\lamEP$ on the real axis as $\epsd \rightarrow \epsEP$.
The original integration contour $\mathcal C$ appears in brown just inside the unit circle.
(b) The $\lambda_\pm$ eigenvalues have coalesced at $\lamEP$, while the contour has been deformed as described in the main text.  A half-pole surrounds $\lamEP$ as evaluated in Eq. (\ref{surv.amp.Q}).
}
\label{fig:contC} 
\end{figure}

We now turn to the evolution on intermediate and long timescales, in which case the EP2A influences the dynamics considerably. For arbitrary time, the survival amplitude in Eq.  (\ref{I.surv.amp}) may be written as
\beq
A(t) = 
\frac{1}{2\pi i} \int_{C_E} dE\, e^{-i Et} \bra d|\frac{1}{E-H_\textrm{I}}|d\ket
=
\frac{1}{2\pi i} \int_{C_E} dE\, e^{-i Et} \bra d|\frac{1}{E-H_{\rm eff}(E)}|d\ket
\label{eq:AHeff}
\eeq
where the contour $C_E$ is shown in Fig. \ref{fig:contCE} and the last equality follows from the perturbation expansion of $(E-H_\textrm{I})^{-1}$ \cite{SHO11}. For our analysis it will be convenient to decompose the Green's function $(E-H_{\rm eff}(E))^{-1}$ into a sum of simpler terms. As shown in \cite{HO14} (see also \cite{TON98}), the generalized eigenvalue problem formalism introduced previously in Sec. \ref{sec:I} gives a decomposition in terms of all the discrete eigenvalues of the Hamiltonian. In App.  \ref{app:ProofSurvAmp} we present the main steps leading to this decomposition, applied specifically to Model I. The final result is 
 \begin{align} \label{eq:xrep}
A(t) 
=\frac{1}{2\pi i}
\sum_{j=\pm} \int_{C}d\lam\,\left(-\lam+\frac{1}{\lam}\right)\exp\left[i\left(\lam+\frac{1}{\lam}\right)t\right]
\bra d|\psi_j\ket\frac{\lam_j}{\lam-\lam_j}\bra\tilde{\psi}_j |d\ket.
\end{align}
Here the summation over $j$ comprises the two discrete eigenstates of the Hamiltonian, which are two virtual bound states to the right of the exceptional point $\epsd \lesssim \epsEP$ or a resonance and anti-resonance pair to the left $\epsd \gtrsim \epsEP$, as discussed beneath Eq. (\ref{I.E.pm.par}).  The integration contour $C$ is shown by the brown, clockwise-oriented circle just inside the unit circle in Fig. \ref{fig:contC}(a).

To evaluate the survival amplitude near the EP2A, we will expand the relevant quantities in terms of 
$\delta$, where $\delta$ quantifies the distance from the exceptional point according to
 \begin{align} \label{eq:ExpEP}
\eps_d = \epsEP + \del
	.
\end{align}
We can then expand the $\lambda$ eigenvalues from Eq. (\ref{I.lambda.pm}) as
 \begin{align} \label{eq:lamrd}
\lam_\pm  = \lamEP\left(1\pm i\del^{1/2}\lamEP^{1/2}- \frac{\lamEP\del}{2}\right) + O(\del^{3/2})
\end{align}
in terms of Eq. (\ref{lambda.EP}).  This type of fractional power eigenvalue expansion (Puiseux expansion) is a universal feature that characterizes the eigenvalues in the vicinity of the exceptional point \cite{Kato,GRHS12,SM_book}.
We also expand the norm of the eigenstates from Eq. (\ref{I.norm})
to obtain
 \begin{align} \label{eq:NormSq}
&\bra d|\psi_\pm\ket\bra{\tilde\psi_\pm}|d\ket =  \bra d|\psi_\pm\ket^2 \approx \frac{1}{2} \left(1\pm \frac{i}{\sqrt{\lamEP\del}}\right)
	.
\end{align}
This last expression serves to quantify the previously noted \footnote{See the comment immediately following Eq. (\ref{I.norm}).} divergence of the norm of the two eigenstates at the exceptional point; however, the divergent terms ultimately cancel between the two contributions when we use this expression to similarly expand the survival amplitude. 

We proceed by putting Eq. (\ref{eq:NormSq}) into Eq. (\ref{eq:xrep}) and expanding to find
 \begin{align} \label{eq:xrep2}
& A(t) \approx \frac{1}{2\pi i} \int_{C}d\lam\,\left(-\lam+\frac{1}{\lam}\right)\exp\left[i\left(\lam+\frac{1}{\lam}\right)t\right]
\left\{  \frac{i}{2\sqrt{\lamEP \del}}
\left[\frac{\lam_+}{\lam-\lam_+} - \frac{\lam_-}{\lam-\lam_-}\right]
+ \frac{1}{2}\left[\frac{\lam_+}{\lam-\lam_+} + \frac{\lam_-}{\lam-\lam_-}\right] \right\}
	.
\end{align}
We next apply the approximation for the $\lambda$ eigenvalues Eq. (\ref{eq:lamrd}) in the integrand of this equation to find
 \begin{align} \label{eq:xrep3}
\frac{\lam_+}{\lam-\lam_+} - \frac{\lam_-}{\lam-\lam_-} \approx  \frac{2 i\lam \lamEP^{3/2} \del^{1/2}}{(\lam-\lamEP)^2},
\end{align}
which gives
 \begin{align} \label{eq:xrep4}
A(t) \approx \frac{1}{2\pi i} \int_{C}d\lam\,\left(-\lam+\frac{1}{\lam}\right)\exp\left[i\left(\lam+\frac{1}{\lam}\right)t\right]
\frac{-\lamEP^2}{(\lam-\lamEP)^2} +O(\del^{1/2})
\end{align}
We see that the divergence in $\delta$ is canceled.

From this point, in the usual case (away from the EP2A) we would simply deform the integration contour in Eq. (\ref{eq:xrep4}) and find there would be two contributions: the residue from the resonance pole and the so-called background integral (to be defined below).  On intermediate timescales the background integral could be ignored in favor of the resonance pole, which would immediately yield the familiar exponential decay evolution.  Meanwhile for the asymptotic timescale, the background integral would instead be the dominant contribution, resulting in inverse power law decay.

In the present case, very near the coalescence at the EP2A, we may deform the contour $C$ according to Fig. \ref{fig:contC}(b).  Here, the contributions from the essential singularities at the origin and at infinity in the upper half plane both vanish.
The remaining contributions to the integral are the half-pole surrounding the coalesced eigenvalue 
$\lamEP$ at the EP2A and the left-running contour along the real axis, which is the background integral.
Taking the half-residue for the $\lamEP$ pole from Eq. (\ref{eq:xrep4}) gives
\beqa
  A_\textrm{P}(t)
     &	= & \frac{1}{2} \left[ 1 + \lamEP^2 + it g^4 \lamEP^3 \right]
		e^{-i \eEP t}
					\nonumber  \\
     &	\approx & \left[ 1 + \frac{1}{4} g^2 + \frac{it g^4}{2} \left( 1 + \frac{3}{2} g^2 \right) \right] e^{-i \eEP t}
	,
\label{surv.amp.Q}
\eeqa
where we have written a further approximation for the small $g$ case in the second line (when the EP2A is close to the band edge).
This result contains the $t e^{-i \eEP t}$ evolution that we would expect by comparison with the EP2B case in the literature \cite{EPexpt1d,HeissSS,Hashimoto1} (although unlike the EP2B case, the purely real eigenvalue $\eEP$ would yield purely non-exponential evolution for the survival probability associated with the pole $P_\textrm{P}(t) = |A_\textrm{P}(t)|^2$).  
However, in practice we find that this result is not only inaccurate, it actually yields non-unitary evolution for all choices of the system parameters and for all values of $t$ (this is true for both the exact result in the first line of Eq. (\ref{surv.amp.Q}) and the approximation in the second line).  Since our model is manifestly Hermitian, any physically sensible result must obey unitary evolution and hence Eq. (\ref{surv.amp.Q}) is useless.

The problem is that when $\eEP$ is very near the continuum threshold, the background integral contribution cannot be neglected even for the intermediate timescale.  However, even when $\eEP$ is not so near the threshold, we find in that case that the short-time parabolic dynamics give a unitary-preserving and much more accurate approximation than Eq. (\ref{surv.amp.Q}).
To accurately capture the influence of the continuum threshold in the former case, we rewrite the expression for the integral Eq. (\ref{eq:xrep4}) entirely, as shown immediately below.


\subsection{Intermediate time evolution when the EP2A is near a band edge}\label{sec:surv.int2}
\begin{figure}
\hspace*{0.05\textwidth}
 \includegraphics[width=0.4\textwidth]{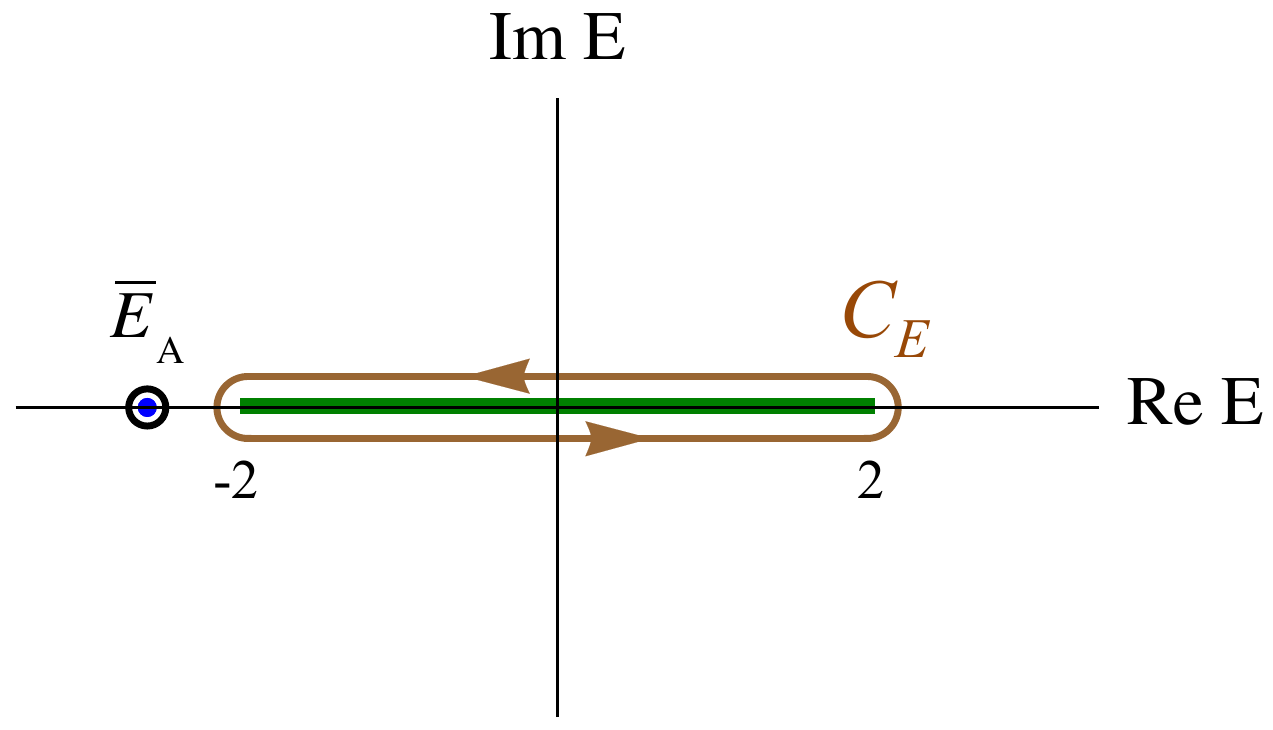}
\hfill
 \includegraphics[width=0.4\textwidth]{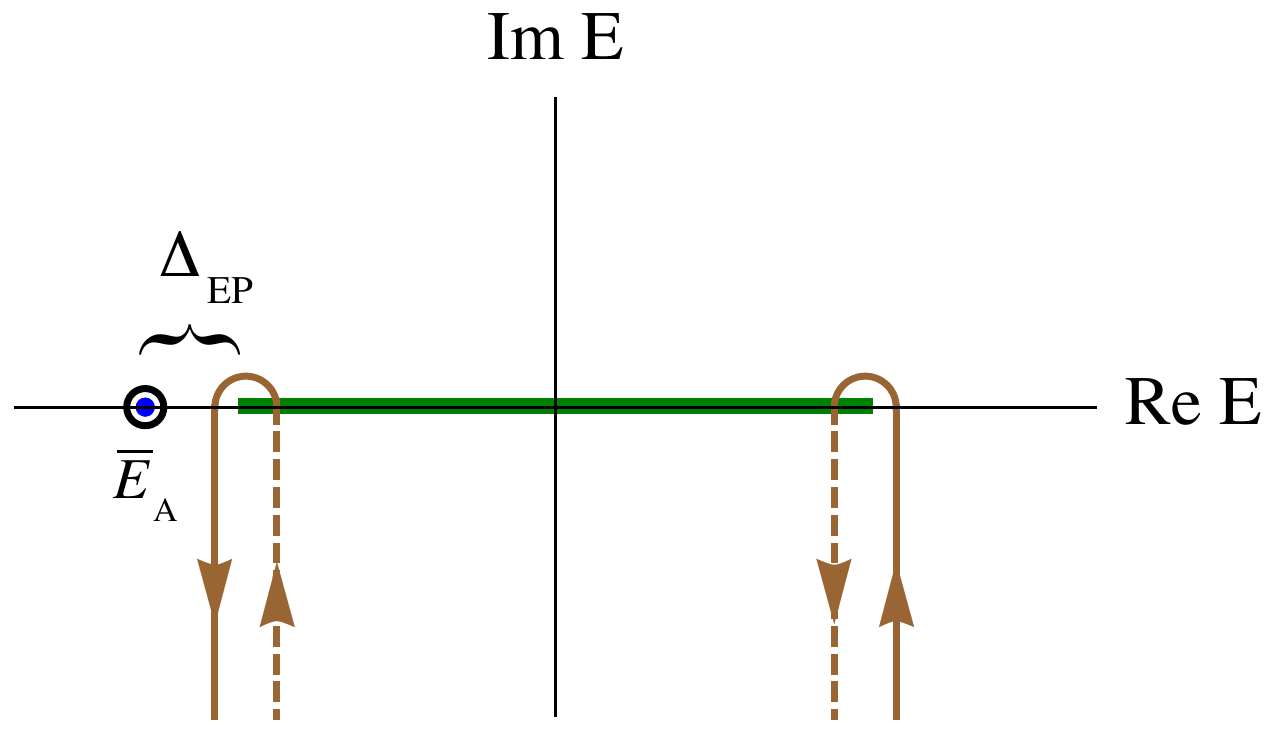}
 \hspace*{0.05\textwidth}
\\
\vspace*{\baselineskip}
\hspace*{0.01\textwidth}(a)\hspace*{0.47\textwidth}(b)\hspace*{0.4\textwidth}
\\
\caption{(color online) Integration contours in the complex $E$ plane.  The continuum extends along the real axis from $E=-2$ to $E=2$, as indicated by the thick green line.  The two level coalescence at the EP2A (which occurs in the second Riemann sheet) is indicated by a blue dot within a black circle.
(a) The contour $C_E$ surrounds the continuum in a counter-clockwise direction in the first sheet.
(b) The deformed contour is shown for the long-time calculation in Eq. (\ref{I.surv.int.bp}); solid brown lines represent portions of the contour appearing in the first sheet, while the dashed brown lines represent portions of the contour in the second sheet.
}
\label{fig:contCE} 
\end{figure}

Going back to Eq. (\ref{eq:xrep4}) the survival amplitude at the EP can  be written as
 \begin{align} \label{eq:Acut2}
& A(t)  
=-\lamEP^2 \frac{\partial \II(\lamEP,t) }{\partial  \lamEP}  
\end{align}
where
 \begin{align} \label{eq:fn}
  \II(\lamEP,t)
=
 \int_{C}\frac{d\lam}{2\pi i}\,\left(-\lam+\frac{1}{\lam}\right)\exp\left[i\left(\lam+\frac{1}{\lam}\right)t\right]
\frac{1}{\lam-\lamEP}.
\end{align}
We find it useful to transform this integral in two stages.  First we transform the integration over the $\lambda$ variable into an integration in the complex energy plane according to the dispersion relation 
$E = - \left( \lambda + 1/\lambda \right)$.  As detailed in App. \ref{appA} [first line of Eq. (\ref{I.int.E.app})], we obtain
 \begin{align} 
\II(\lamEP,t) 
=
- \frac{1}{2 \pi} \int_{C_E} dE\,e^{-i E t} 
\frac{\sqrt{1-E^2/4}}{ E-\eEP}
	,
\label{I.int.E}
\end{align}
where the contour $C_E$ is shown in Fig. \ref{fig:contCE}(a).
This expression for the integral will be useful when we analyze the long-time dynamics later on.  However, for our immediate purposes it is more useful to further rewrite this expression to obtain the final form of the integral from App. \ref{appA}, given by
 \begin{align} \label{eq:fn2}
 \II(\lamEP,t)
= e^{-i \eEP t}\left[ 1/\lamEP -i  \int_{0} ^t dt' \, e^{i \eEP t'}\frac{J_1(2t')}{t'} \right]
\end{align}
in which $\eEP=-\lamEP-\lamEP^{-1}$; see Eq. (\ref{E.EP}).

Inserting this last expression into Eq. (\ref{eq:Acut2}) we obtain
 \begin{align} \label{eq:fn3}
 A(t)
= e^{-i \eEP t}\left[1 + \frac{\partial \eEP}{\partial \lamEP} \left(it \lamEP +  \lamEP^2  \int_{0} ^t dt' \, e^{i \eEP t'} J_1(2t')\left( \frac{t}{t'}-1\right) \right)\right]
	.
\end{align}
Then, using the expansion Eq. (\ref{eq:I10}) we get 
 \begin{align} \label{eq:fn4}
 A(t)
= e^{-i \eEP t}\left[1 + \frac{\partial \eEP}{\partial \lamEP} \left(it \lamEP +  \lamEP^2  \sum_{l=0}^\infty \frac{(-i\delEP)^l}{l!} \left(t K_l(t) -K_{l+1}(t)\right)\right)\right]
\end{align}
in which the functions $K_l(t)$ are integrals of Bessel's functions [see Eqs. (\ref{eq:K0}-\ref{eq:K22})] and $\delEP$ is defined as
\beq
  \delEP
  	\equiv -2-\eEP,
\label{del.ep}
\eeq
which is the distance between the EP2A energy eigenvalue 
$\eEP$ and the band edge at $E=-2$ (note that $\delEP$  is a positive quantity since $\eEP < -2$, although we sometimes use absolute value bars below to make this explicit).

Assuming $|\delEP t|\ll 1$, or $t \ll T_\textrm{EP}$ with
\beq
  T_\textrm{EP}
  	\equiv \frac{1}{| \delEP |},
\label{T.EP}
\eeq
we keep only terms of order $(\delEP t)^n$ with $n=1/2$ or $n=1$, which implies keeping only the $l=0$ term in Eq. (\ref{eq:fn4}). Moreover, using the approximations  (for $z\ge 2$)
 \begin{align} \label{eq:BAsymp}
& J_0(z) \approx \sqrt{\frac{2}{\pi z}} \cos\left(z-\pi/4\right) \nonumber\\
& J_1(z) \approx \sqrt{\frac{2}{\pi z}} \cos\left(z-\pi/4 - \pi/2\right)
\end{align}
we obtain
 \begin{align} \label{eq:BAsymp2}
A_\Delta (t) \approx e^{-i \eEP t}\left[1 - 4 \sqrt{\frac{ i t\delEP }{\pi}} + 2 it \delEP\right] + O(t\delEP )^{3/2}
\end{align}
This gives the intermediate-time evolution when the EP is very near the lower band edge. Note that since we assumed $|t\delEP  |\ll 1$ to obtain this result, the $\sqrt{t}$ term is much larger than the $t$ term inside the brackets. 

In order to better understand the origin of the $\sqrt{t}$ term in Eq. (\ref{eq:BAsymp2}), note that the condition $|t\delEP  |\ll 1$ is essentially equivalent to that for the long time near zone discussed in Sec. \ref{sec:surv.cont}.  Hence, the $t e^{-i \eEP t}$ term mentioned below Eq. (\ref{surv.amp.Q}), which we would expect in the survival amplitude from comparison with the EP2B case, has been modified by a $t^{-1/2}$ factor from proximity to the continuum threshold to give $t^{-1/2} * t e^{-i \eEP t} \sim t^{1/2} e^{-i \eEP t}$.

\begin{figure}
\hspace*{0.05\textwidth}
 \includegraphics[width=0.4\textwidth]{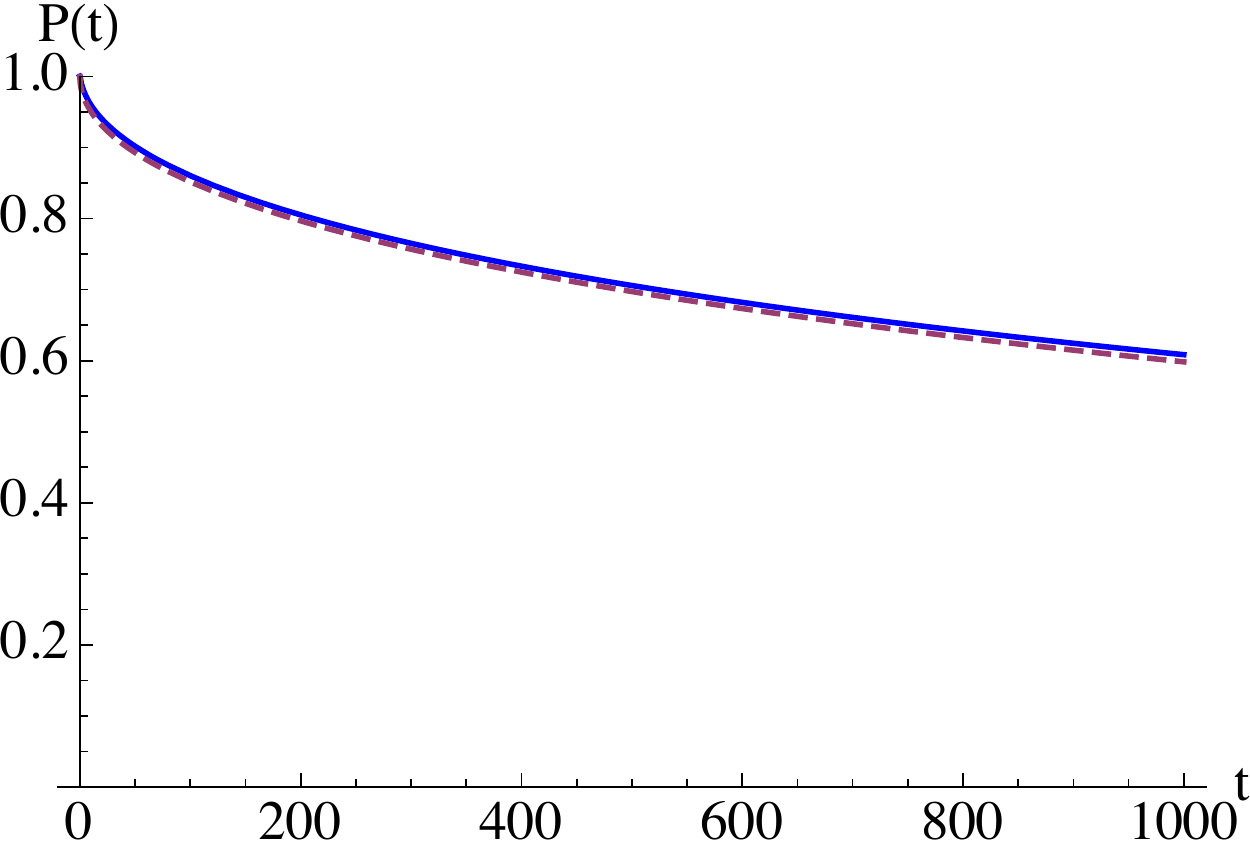}
\hfill
 \includegraphics[width=0.4\textwidth]{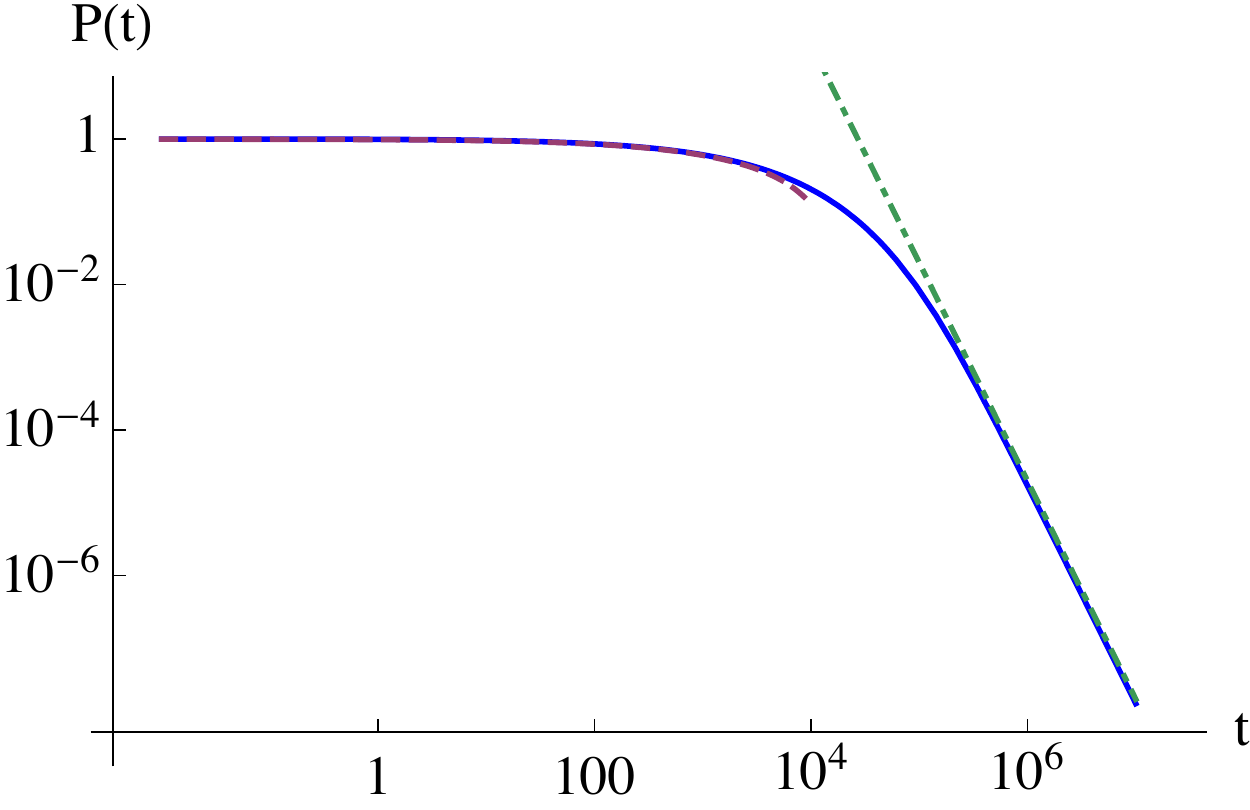}
 \hspace*{0.05\textwidth}
\\
\vspace*{\baselineskip}
\hspace*{0.01\textwidth}(a)\hspace*{0.47\textwidth}(b)\hspace*{0.4\textwidth}
\\
\caption{(color online) (a) Linear plot and (b) log-log plot of the survival probability $P(t)$ near to the EP2A, which in turn is near the band edge.  The solid blue curve gives the numerical integration of Eq. (\ref{eq:xrep}) while the purple dashed  curve represents the approximate expression for the survival probability $P_\Delta (t)$ reported in Eq. (\ref{I.P.band.edge}).
Here $g=0.1$, which places the EP2A in the energy plane at $\eEP \approx -2.0000253$, very near the band edge at $E=-2$, and in the parameter space at $\epsEP \approx -1.989975$.  We then choose $\epsd = -1.989974$, near the EP. In (b), the green chained line is the long-time approximation $P_\textrm{LT} (t)$ in Eq. (\ref{I.surv.prob.fz}).
 }
\label{fig:P.band.edge}
\end{figure}

Finally, we take the square modulus of Eq. (\ref{eq:BAsymp2}) to obtain the survival probability when the EP2A is near to the band edge as
\beq
  P_\Delta (t)
  	\approx 1 -4 \sqrt{ \frac{2 t |\delEP|}{\pi}} + \frac{16 t |\delEP|}{\pi}
	.
\label{I.P.band.edge}
\eeq
This result is plotted (dashed curve) against the numerical integration (solid curve) in Fig. \ref{fig:P.band.edge} (a,b) for the case $g=0.1$ and $\epsd = -1.989974$.  With these values, the EP is located in the energy plane at $\eEP \approx -2.0000253$, very near the band edge at $E=-2$.  In the parameter space, 
$\epsd = -1.989974$ is slightly shifted from the EP at $\epsEP \approx -1.989975$.  In this case, Eq. (\ref{I.P.band.edge}) describes the evolution quite accurately up until about 
$t \gtrsim T_\textrm{EP}$, where $T_\textrm{EP} = 1/|\delEP|  \sim 10^5$ as defined in Eq. (\ref{T.EP}).  
For $t \gg T_\textrm{EP}$,
 the survival probability follows the long time inverse power law evolution $P(t) \sim t^{-3}$ (chained line) as obtained next in Sec. \ref{sec:surv.long}.

\begin{figure}
\hspace*{0.05\textwidth}
 \includegraphics[width=0.4\textwidth]{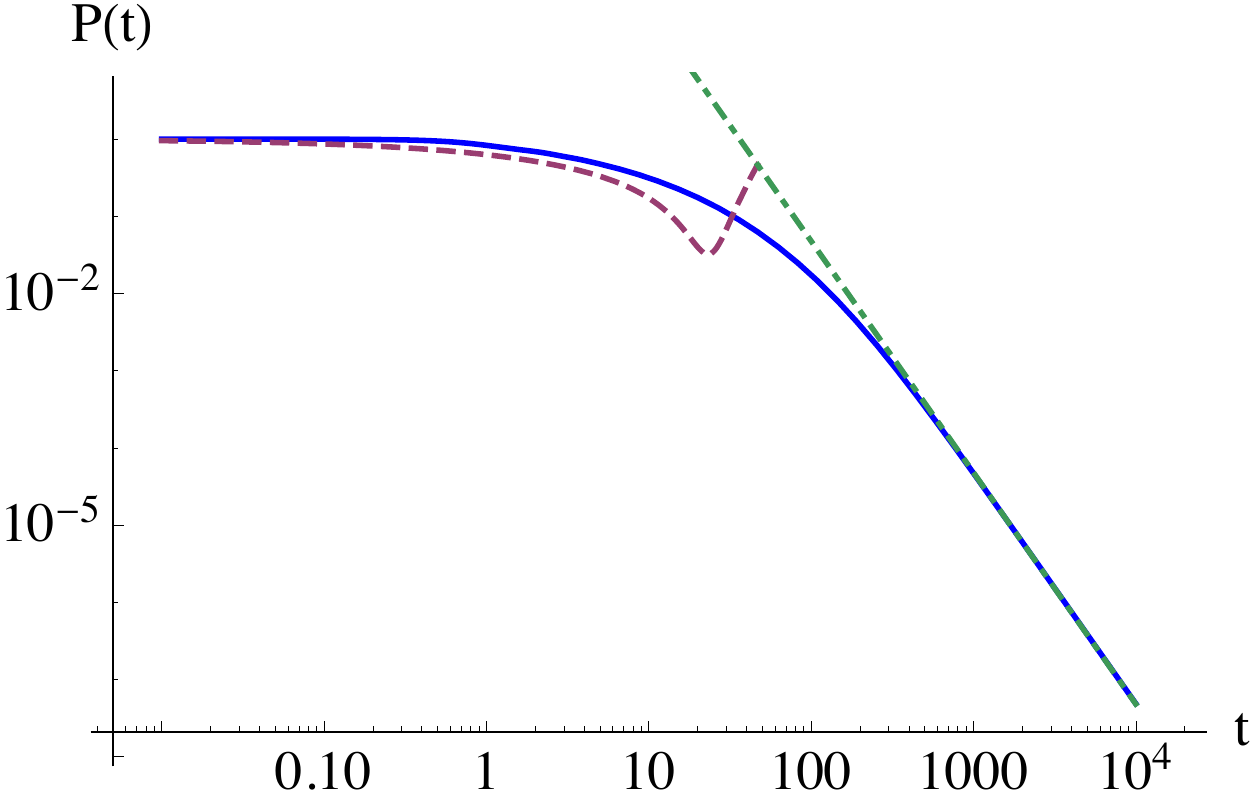}
\hfill
 \includegraphics[width=0.4\textwidth]{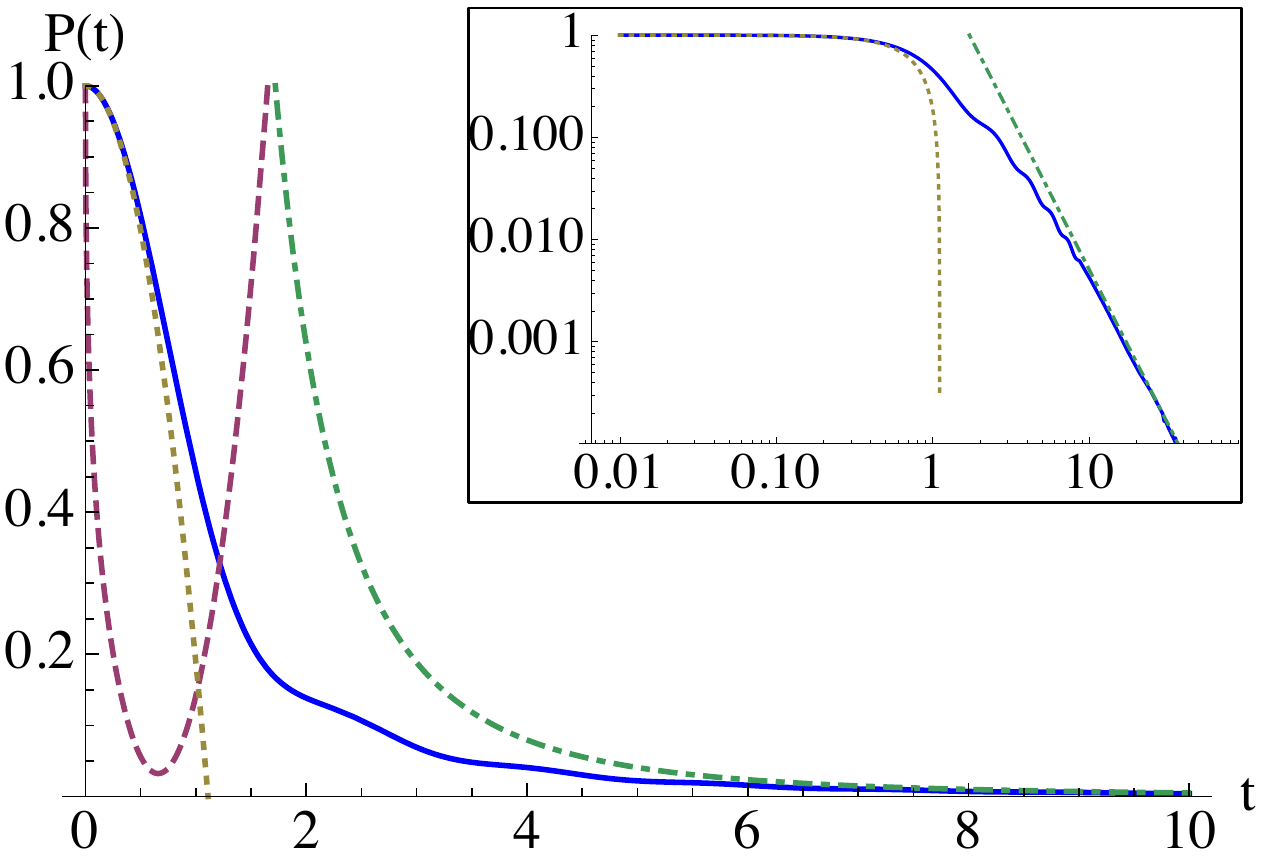}
 \hspace*{0.05\textwidth}
\\
\vspace*{\baselineskip}
\hspace*{0.01\textwidth}(a)\hspace*{0.47\textwidth}(b)\hspace*{0.4\textwidth}
\\
\caption{(color online) (a) Log-log plot of the survival probability $P(t)$ near the EP2A 
for $\eEP \approx -2.02073$ (with $g=0.5$), which is somewhat farther from the band edge compared to Fig. \ref{fig:P.band.edge}.  The solid blue curve gives the numerical integration of Eq. (\ref{eq:xrep}), the dashed purple curve represents the expression for the survival probability $P_\Delta (t)$ reported in Eq. (\ref{I.P.band.edge}) and the chained green line gives the long time evolution $\sim t^{-3}$ from  Eq. (\ref{I.surv.prob.fz}). For $g=0.5$ the EP2A is located in the parameter space at $\epsEP \approx -1.73205$; we have then choosen $\epsd = -1.7321$.
(b) Linear plot of the survival probability near to the EP2A when it is even farther from the band edge [inset: log-log plot].  Here $g=0.9$, which places the EP2A in the energy plane at $\eEP \approx -2.73005$ (somewhat far from the band edge) and at $\epsEP \approx -0.87178$ in the parameter space. We have set $\epsd = -0.87$.
In (b), the approximation $P_\Delta (t)$ in Eq. (\ref{I.P.band.edge}) (purple dashed curve) fails because the EP2A is too far from the band edge. In this situation, the short-time parabolic dynamics $P_\textrm{Z}(t) \approx 1 - g^2 t^2$ in Eq. (\ref{I.surv.prob.zeno}) (beige dotted curve) are strongly pronounced. }
\label{fig:P.band.edge.2}
\end{figure}

A similar evolution occurs for the case $g=0.5$ in Fig. \ref{fig:P.band.edge.2}(a), although the EP2A is now located a bit further from the threshold at $\eEP \approx -2.02073$.
Because the gap between the EP2A and the threshold has increased, the timescale 
$T_\textrm{EP} \sim 50$ at which the system begins the transition to the far zone dynamics occurs much earlier.

In the case $g=0.9$ with $\eEP \approx -2.73005$ in Fig. \ref{fig:P.band.edge.2}(b) the timescale $T_\textrm{EP}$ is further decreased and now falls approximately at the timescale 
$T_\textrm{Z} \approx 1.15$ [from Eq. (\ref{T.Z})]
associated with the short-time dynamics.  
As a result, the system decays relatively very quickly, mostly during the time region $t \lesssim T_\textrm{Z}$ in which the parabolic dynamics $1 - g^2 t^2$ from Eq. (\ref{I.surv.prob.zeno}) dominates and hence this effect is strongly enhanced.  Note that this is related to the enhanced short-time dynamics appearing in Ref. \cite{Longhi06}, although for somewhat different model parameters in that case.

\subsection{Inverse power law evolution on long timescales near the EP2A}\label{sec:surv.long}

It is well-established that on long time scales the asymptotic dynamics in typical quantum systems follows an inverse power law decay \cite{Fonda,Muga_review} associated with the existence of the continuum threshold (band edge) \cite{Khalfin,Hack}.  Starting with the expression for the integral $\II(\lamEP,t)$ reported in Eq. (\ref{I.int.E}), we can obtain a useful approximation for the survival probability in this time region by deforming the contour integral in the complex energy plane as follows.

The integration in Eq. (\ref{I.int.E}) is presently written in terms of a counter-clockwise contour surrounding the branch cut in the first Riemann sheet of the complex energy plane, as shown in Fig. \ref{fig:contCE}(a).  We can deform this contour by dragging both horizontal segments out to infinity in the lower half plane.  As the contour originally located in the upper half plane is dragged into the lower half plane it passes through the branch cut extending from $-2$ to $2$ and crosses into the second Riemann sheet.  Note that, owing to the factor $e^{-iEt}$ in Eq. (\ref{I.int.E}), the integration associated with the portions of the contour at infinity in the lower half plane vanish; the remaining non-vanishing contributions of the integral are shown in Fig. \ref{fig:contCE}(b) and given by
\beq
  \II(\lamEP,t)
  	= \II_\textrm{L} (\lamEP,t) + \II_\textrm{U} (\lamEP,t)
\eeq
where $\II_\textrm{L(U)} (\lamEP,t)$ gives the contribution remaining from the lower (upper) band edge, written as
\beqa
  \II_\textrm{X} (\lamEP,t)
    &	= & \frac{1}{2 \pi} \left[  
			\int_{\mp 2 - i \infty}^{\mp 2} dE e^{-iEt} \frac{\sqrt{1 - E^2/4}}{E - \eEP}
				- \int_{\mp 2}^{\mp 2 - i \infty} dE e^{-iEt} \frac{\sqrt{1 - E^2/4}}{E - \eEP}
		\right]
							\nonumber \\
   &	= & - \frac{1}{\pi} \int_{\mp 2}^{\mp 2 - i \infty} dE e^{-iEt} \frac{\sqrt{1 - E^2/4}}{E - \eEP}
\label{I.surv.int.bp}
\eeqa
where the $\textrm{X}=\textrm{L}(\textrm{U})$ contribution is given by the upper (lower) sign.  

Assuming the two eigenvalues appear in the vicinity of (or below) the lower band edge, then we would expect that the $\textrm{X}=\textrm{L}$ integral would be the dominant contribution and the $\textrm{X}=\textrm{U}$ integral can safely be neglected, which is borne out in the numerics.  To estimate the remaining term, we first rewrite the integration in a more natural way according to $E = -2 - iy$, in which $y \in [0,\infty]$.  Thus we obtain
\beq
  \II_\textrm{L} (\lamEP,t) 
  	= \frac{i e^{2it}}{2\pi} 
			\int_0^\infty dy \; e^{-yt} \frac{\sqrt{y^2 - 4 i y}}{\Delta_\textrm{EP} - i y}
	,
\label{I.surv.int.bp.y}
\eeq
where we have used the definition for $\delEP$ from Eq. (\ref{del.ep}). 
We next rewrite the integration variable again as $s = -yt$ to obtain
\beqa
  \II_\textrm{L} (\lamEP,t) 
     &	= & \frac{i e^{2it}}{2\pi \Delta_\textrm{EP} t^2} 
			\int_0^\infty ds \; e^{-s} \frac{\sqrt{s^2 - 4 i s t}}{1 + s \frac{1}{i t \Delta_\textrm{EP}}}
	,								\nonumber \\
     &	\approx & \frac{i e^{2it}}{2\pi \Delta_\textrm{EP} t^2} 
			\int_0^\infty ds \; e^{-s} \sqrt{s^2 - 4 i s t}
\label{I.surv.int.bp.s}
\eeqa
where in the second line we have taken the long time approximation 
$t \gg T_\textrm{EP}$
from Eq. (\ref{T.EP}).
The remaining integration can be approximated as 
\beqa
\int_0^\infty ds \; e^{-s} \sqrt{s^2 - 4 i s t}  \approx  \sqrt{-4it} \int_0^\infty ds \; \sqrt{s} e^{-s} = \sqrt{\pi t} e^{-i\pi/4},
\label{g_int}
\eeqa
which gives
\beq
  \II_\textrm{L} (\lamEP,t) 
	= \frac{e^{i \pi / 4}e^{2it}}{2 \sqrt{\pi} \Delta_\textrm{EP} t^{3/2}}
	.
\label{I.surv.int.bp.final}
\eeq
Putting this result in Eq. (\ref{eq:Acut2}), taking the appropriate derivative and re-arranging gives the long-time contribution to 
the survival probability as
\beq
  P_\textrm{LT}(t)
	\approx \frac{g^4}{4\pi \left( 1 - \sqrt{1-g^2} \right)^8 t^{3}}
	.
\label{I.surv.prob.fz}
\eeq
This result describes the system dynamics near the EP on the longest time scales,
for which the $t^3$ effect is quite typical in open quantum systems \cite{GCV06,GPSS13,GCMM95,MDS95,Zueco16,DBP08,HO14}.  Hence we see that while the EP dramatically modifies the dynamics on intermediate time scales, the long time dynamics are qualitatively the same as the ordinary case away from the EP; although the time scale $T_\textrm{EP}$ at which the long time effect sets in may be modified according to the proximity of the EP2A to the band edge.


\section{Model II: Time evolution at the EP2B}\label{sec:II.ham}

In this section we briefly demonstrate the time evolution near the EP2B appearing in Model II.
First we derive the spectrum and the mapping to the generalized eigenvalue problem just as we did for Model I; however, here we only give a quick outline of the development as the details are a fairly straightforward generalization from the Model I case.  We emphasize our primary objective here is to verify in the present formalism the modified exponential decay on intermediate timescales due to the EP2B \cite{EPexpt1d} and demonstrate that the power law decay resulting from the continuum threshold still takes hold asymptotically.

First we can write the Hamiltonian for Model II from Eq. (\ref{ham.model.II.intro}) in the single particle picture as
\beq
  H_\textrm{II}
  	= -V \left( | \dA \ket \bra \dB | +  | \dB \ket \bra \dA |  \right) 
		- g \left( | c_1 \ket \bra \dB | + | \dB \ket \bra c_1 | \right)
			- \sum_{j=1}^{\infty} \left( | j \ket \bra j+1 | + | j+1 \ket \bra j | \right)
	,
\label{ham.model.II}
\eeq
where again we set the energy units as $b=1$ and assume $g < 1$.
Then we write the wave function as an outgoing wave from the double impurity region as
\begin{equation}
  \psi (x) =
	\left\{ \begin{array}{ll}
		\psi_\textrm{A}				& \mbox{for $x = \textrm{A}$}	\\
		\psi_\textrm{B}				& \mbox{for $x = \textrm{B}$}   	\\
		C e^{ikx}					& \mbox{for $x \ge 1$}
	\end{array}
	\right.  
	.
\label{II.outgoing}
\end{equation}
It is easy to again show that the dispersion in the leads follows $E = - 2 \cos k$.

We next write the Schr\"odinger equation for the wave function in Eq. (\ref{II.outgoing}) and follow a method similar to Sec. \ref{sec:I} to obtain the eigenvalue equation in terms of the effective Hamiltonian as
\begin{equation}
\left( 
\begin{array}{cc}
	0	& -V				\\
	-V	& -g^2 e^{ik_j}	\\ 
		\end{array}
	\right)
\left(  \begin{array}{c}
		\bra \dA | \psi_j \ket  	\\
		\bra \dB | \psi_j \ket  
		\end{array}
	\right)
= E_j
\left(  \begin{array}{c}
		\bra \dA | \psi_j \ket  	\\
		\bra \dB | \psi_j \ket 
		\end{array}
	\right) ,
\label{II.H.eff.eqn}
\end{equation}
which is just the realization of Eq. (\ref{H.eff.schr}) for Model II.
Taking the determinant of this equation and again using the tight-binding dispersion $E_j = - 2 \cos k_j$ reveals the polynomial for the discrete eigenvalues $p_\textrm{II} (E_j) = 0$ as a quartic given by
\begin{equation}
  p_\textrm{II}(E)
  	= \left( 1 - g^2 \right) E^4
		+ \left[ g^4 + \left( g^2 - 2 \right) V^2 \right] E^2 + V^4
	.
\label{II.disp.E}
\end{equation}
The four energy eigenvalues found from this equation are given by all four sign combinations in
\begin{equation}
  E_j
  	= \pm \sqrt{
		\frac{ \left( 2 - g^2 \right) V^2 - g^4 \pm g^2 \sqrt{g^4 + 2 \left( g^2 - 2 \right)V^2 + V^4}}
			{2 \left( 1 - g^2 \right)} }
	.
\label{II.E.j}
\end{equation}

We can immediately find the exceptional points as the zeroes of the inner square root in Eq. (\ref{II.E.j}).  Let us assume that $g < 1$ is a fixed parameter while $V$ can be externally controlled.  Then solving 
$g^4 + 2 \left( g^2 - 2 \right)V^2 + V^4 = 0$ yields
\begin{equation}
  \bar{V}_\textrm{L} 
  	= 1 \pm \sqrt{1-g^2}
	.
\label{past.EP.V.j}
\end{equation}
as the location of the EPs in parameter space.  By plugging these values back into Eq. (\ref{II.E.j}) we can further show that the upper sign gives an EP2A while the lower sign gives an EP2B.  Since our primary interest in Model II is the EP2B, we focus on this case from this point forward.

It can be shown that immediately on either side of the EP2B, the spectrum consists of two resonance states and two anti-resonance states.  Directly at the EP at $V = \VB$, the resonances coalesce and the two anti-resonances coalesce on a pure imaginary eigenvalue $E = \pm \eB$.  The location and eigenvalues of the EP2B are given by
\begin{equation}
  \VB 
  	= 1 - \sqrt{1-g^2} \ ;
					\ \ \ \ \ \ \ \ \ \ \ \ \ \ \ \ \ \ \ 
  \pm \eB 
  	= \pm i \sqrt{\frac{2-g^2}{\sqrt{1-g^2}} - 2}
	 \equiv \pm i \frac{\gamB}{2}
		,
\label{II.EP2B}
\end{equation}
where we have defined the decay width at the EP as $\gamB$.


\subsection{Mapping to the generalized eigenvalue problem for Model II}\label{sec:II.GEP}

We now transform the non-standard eigenvalue problem Eq. (\ref{II.H.eff.eqn}) for Model II into a generalized linear eigenvalue problem, following a similar development as presented in Sec. \ref{sec:models.GEP}.  First we again introduce the variable $\lambda \equiv e^{ik}$, which allows us to write Eq. (\ref{II.H.eff.eqn}) as a 2x2 quadratic eigenvalue problem.  This quadratic eigenvalue problem is then in turn transformed into a 4x4 generalized linear eigenvalue problem of the form $(F - \lambda G) | \Psi \ket = 0$ with
\beq
  F = 
	\left[ 
	\begin{array}{cccc}
		0	& 0	& 1		& 0				\\
		0	& 0	& 0		& 1				\\
		1 	& 0	& 0		& -V				\\
		0	& 1	& -V		& 0
			\end{array}
		\right]
			\ \ \ \ \ \ \ \ \ \ \ \ \ \ \ \ \ \ \ 
  G = 
	\left[ 
	\begin{array}{cccc}
		1	& 0	& 0		& 0				\\
		0	& 1	& 0		& 0				\\
		0 	& 0	& -1		& 0				\\
		0	& 0	& 0		& g^2 - 1
			\end{array}
		\right]
	.
\label{II.A.B}
\eeq
and
\beq
  | \Psi \ket
	= 	\left[ 
	\begin{array}{c}
		\bra \dA | \psi \ket	\\
		\bra \dB | \psi \ket	\\
		\lambda \bra \dA | \psi \ket 	\\
		\lambda \bra \dB | \psi \ket
			\end{array}
		\right]
	.
\label{II.Psi.defn}
\eeq
By taking the determinant of the eigenvalue equation above we obtain the four $\lambda$ eigenvalues as
\begin{equation}
  \lambda_{j}
  	= \pm \sqrt{
		\frac{ V^2 + g^2 - 2 \pm \sqrt{g^4 + 2 \left( g^2 - 2 \right)V^2 + V^4}}
			{2 \left( 1 - g^2 \right)} }
	.
\label{II.lambda.j}
\end{equation}
We can plug in $\bar{V}_\textrm{B} = 1 - \sqrt{1-g^2}$ from Eq. (\ref{II.EP2B}) to obtain the values of 
$\lambda_j$ at the EP2B as
\begin{equation}
  \pm \bar{\lambda}_\textrm{B} 
  	= \pm i \frac{1}{\left( 1-g^2 \right)^{1/4}}
.
\label{II.lambda.B}
\end{equation}
We emphasize that here the $+$ sign choice corresponds to the energy where the two resonances coalesce (i.e., $-\lamB - 1/\lamB = - \eB = - i \gamB$), while the $-$ sign corresponds to that at which the anti-resonances coalesce.  Next, we expand in the parameter space in the vicinity of the EP2B according to
\begin{equation}
  V 
  	= \VB + \delta
	.
\label{II.V.exp}
\end{equation}
Then the four $\lambda$ eigenvalues from Eq. (\ref{II.lambda.j}) can be expanded 
as
\begin{eqnarray}
  \lambda_{s,\pm}
     &	\approx & s \frac{i}{\left( 1 - g^2 \right)^{1/4}} 
     				\left[ 1 \mp \sqrt{\frac{1}{2} \left( 1 - \frac{1}{\sqrt{1-g^2}} \right)} \delta^{1/2} 
				+ \frac{1}{4} \left( 1 - \frac{1}{\sqrt{1 - g^2}} \right) \delta
				\mp \frac{ \delta^{3/2} }{ 4\sqrt{2} \sqrt{1 - g^2 - \sqrt{1-g^2}} }
				+ O (\delta^2)
				\right]
	,
\label{II.lambda.j.exp}
\end{eqnarray}
where $s=+$ ($s=-$) gives the expansion for the two coalescing resonances (anti-resonances).

In the calculation in Sec. \ref{sec:II.surv} below, we will consider the survival probability of the initially occupied $| \dA \ket$ state.  Hence, we also need an explicit expression for this eigenvector.  To obtain this, we first write the eigenket equation $(F - \lambda_j G) | \Psi \ket = 0$ in explicit form as
\begin{equation}
	\left( 
	\begin{array}{cccc}
		- \lambda_{j}	& 0			& 1			& 0	\\
		0			& -\lambda_{j}	& 0			& 1	\\
		1 			& 0			& \lambda_{j}	& - V	\\
		0			& 1			& -V			& \left( 1-g^2 \right) \lambda_{j}
			\end{array}
		\right)
	\left( 
	\begin{array}{c}
		\bra \dA | \psi \ket	\\
		\bra \dB | \psi \ket	\\
		\lambda_j \bra \dA | \psi \ket 	\\
		\lambda_j \bra \dB | \psi \ket
			\end{array}
		\right)
		= 0
	,
\label{II.GEP.eigenket.eqn}
\end{equation}
where we have used Eqs. (\ref{II.A.B}) and (\ref{II.Psi.defn}).
Combining the third line of this matrix equation with the normalization condition $\bra \tilde{\Psi}_j | B | \Psi_j \ket = 1$, we obtain the contribution of the $\dA$ site to the eigenket as
\begin{equation}
  \bra \dA | \psi_j \ket^2
  	= \frac{V^2 \lambda_j^2}
		{1 + \left( 1 + g^2 + V^2 \right) \lambda_j^2 
			- \left( 1 - 2g^2 + V^2 \right) \lambda_j^4 - \left( 1 - g^2 \right) \lambda_j^6}
		.
\label{II.GEP.eigenket.dA}
\end{equation}
We can again expand this quantity in terms of Eq. (\ref{II.V.exp}), in which case we obtain
\begin{eqnarray}
	\bra \dA | \psi_{s,\pm} \ket^2
  	= \frac{1}{4} \left[ 1 \mp \frac{1}{\sqrt{2 \delta}} \sqrt{1 - g^2 - \sqrt{1-g^2}} 
		\mp \frac{ \left(1 - 3g^2 + \left( 1 - g^2 \right)^{3/2} \right)  \sqrt{1 - g^2 - \sqrt{1-g^2}} }
			{4 \sqrt{2} g^2} 												          \delta^{1/2}
	+ O(\delta)
			\right]
		.
\label{II.GEP.eigenket.dA.exp}
\end{eqnarray}
Notice there is no $s$-dependence ($s=\pm$) for this squared component.  Also note that for compactness we do not report the $\delta$-order term in this expression, although we will use that term in the calculation of the survival amplitude below in order to maintain consistency.


\subsection{Time evolution near the EP2B}\label{sec:II.surv}

We now turn to the time evolution near the EP2B in Model II. We assume the $| \dA \ket$ state is initially occupied at $t = 0$ for definiteness, although the results would be qualitatively similar for the $| \dB \ket$ state or some linear combination of the two. 

First we can again easily calculate the parabolic evolution on short time scales for $P(t) = |A(t)|^2 =  | \bra \dA | e^{-iHt} | \dA \ket |^2$  using Eq. (\ref{I.surv.amp.zeno.H}), except substituting the Hamiltonian for Model II from Eq. (\ref{ham.model.II}).  Doing so we immediately find
\beq
P(t)
	\approx 1 - V^2 t^2
	,
\label{II.surv.amp.zeno}
\eeq
which holds for $t \ll 1/\sqrt{2} V$.
We again emphasize this result holds on this timescale either nearby or far away from any exceptional point in Model II. 

\begin{figure}
\hspace*{0.05\textwidth}
 \includegraphics[width=0.4\textwidth]{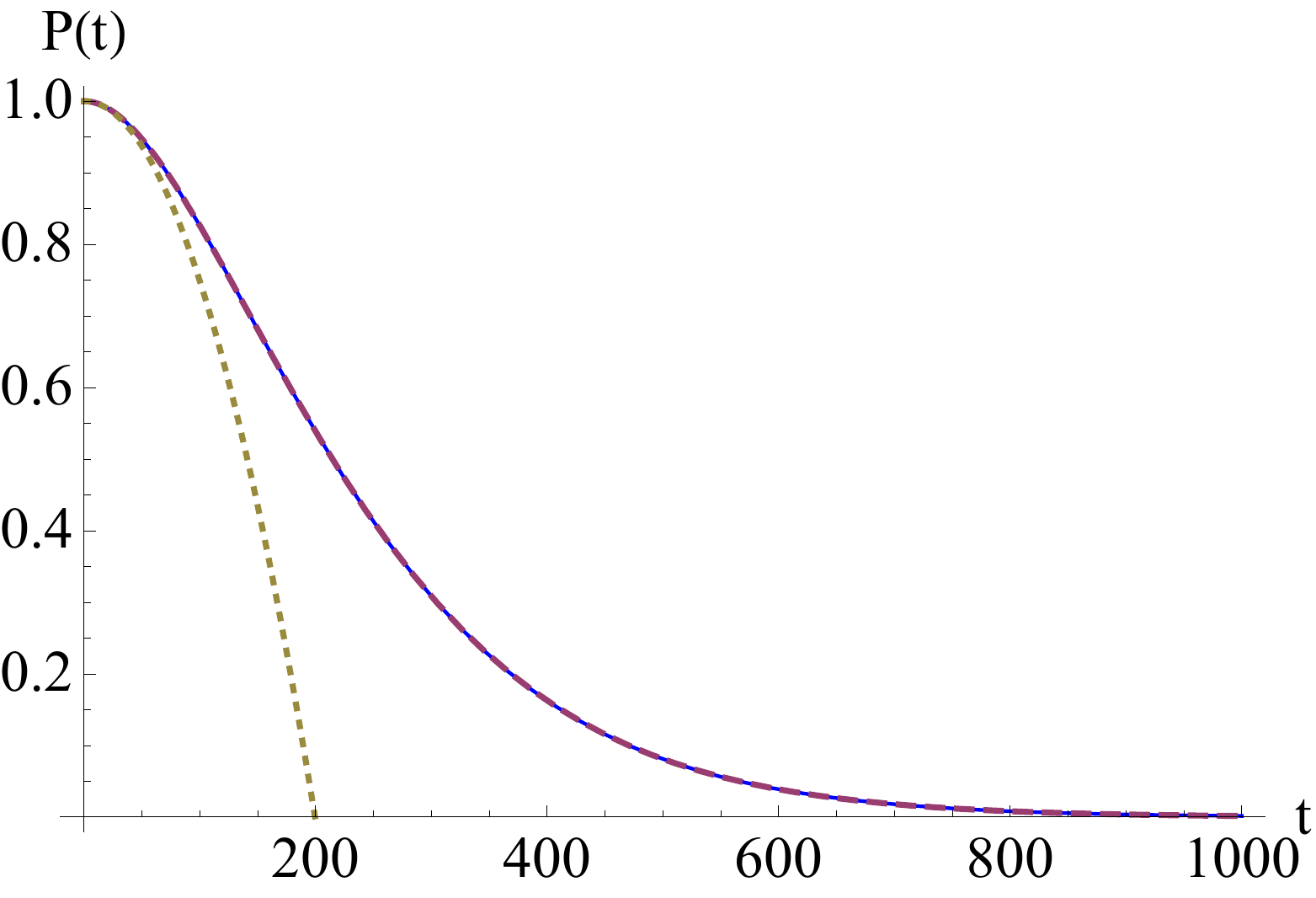}
\hfill
 \includegraphics[width=0.4\textwidth]{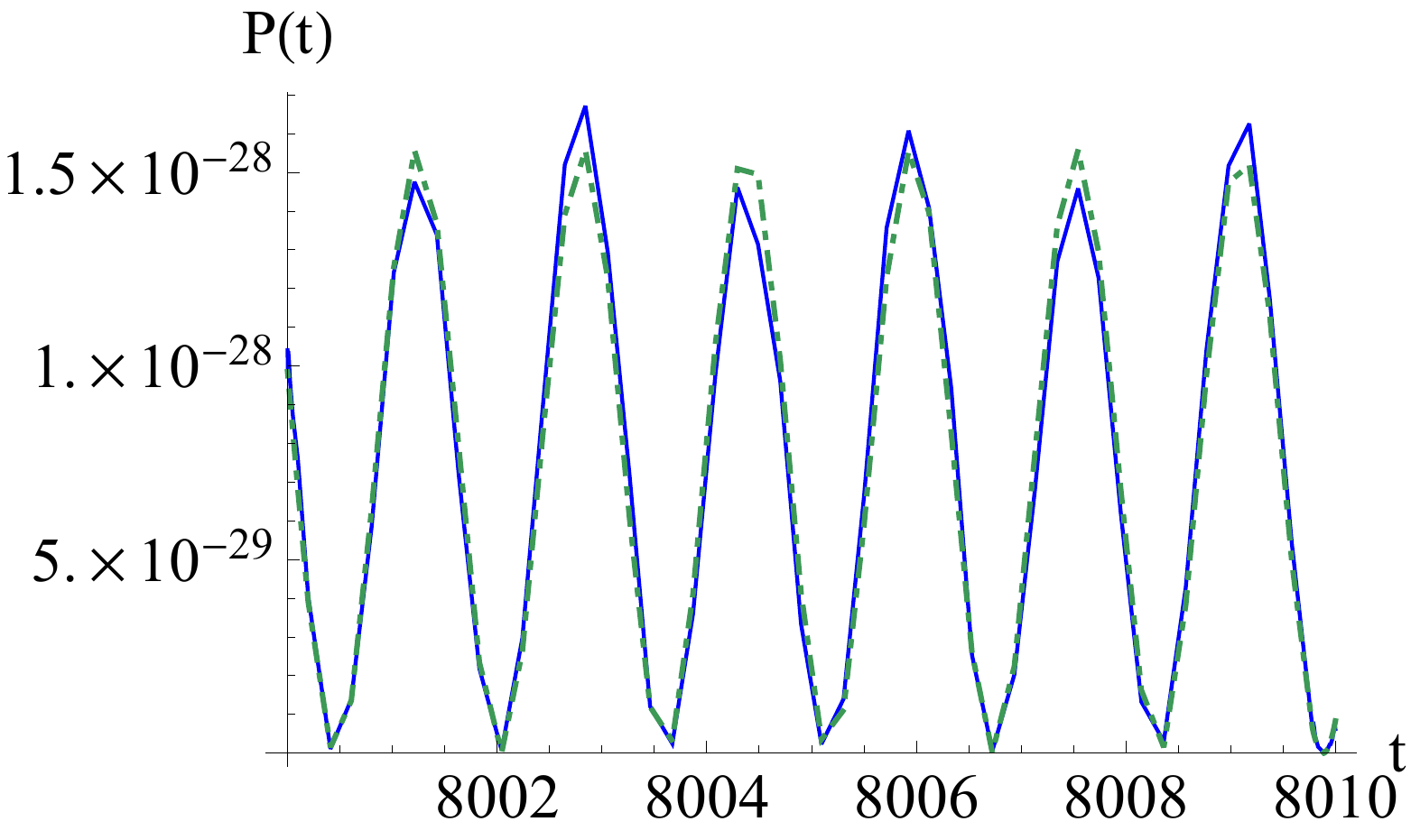}
 \hspace*{0.05\textwidth}
\\
\hspace*{0.01\textwidth}(a)\hspace*{0.47\textwidth}(b)\hspace*{0.4\textwidth}
\\
\vspace*{\baselineskip}
\hspace*{0.05\textwidth}
 \includegraphics[width=0.4\textwidth]{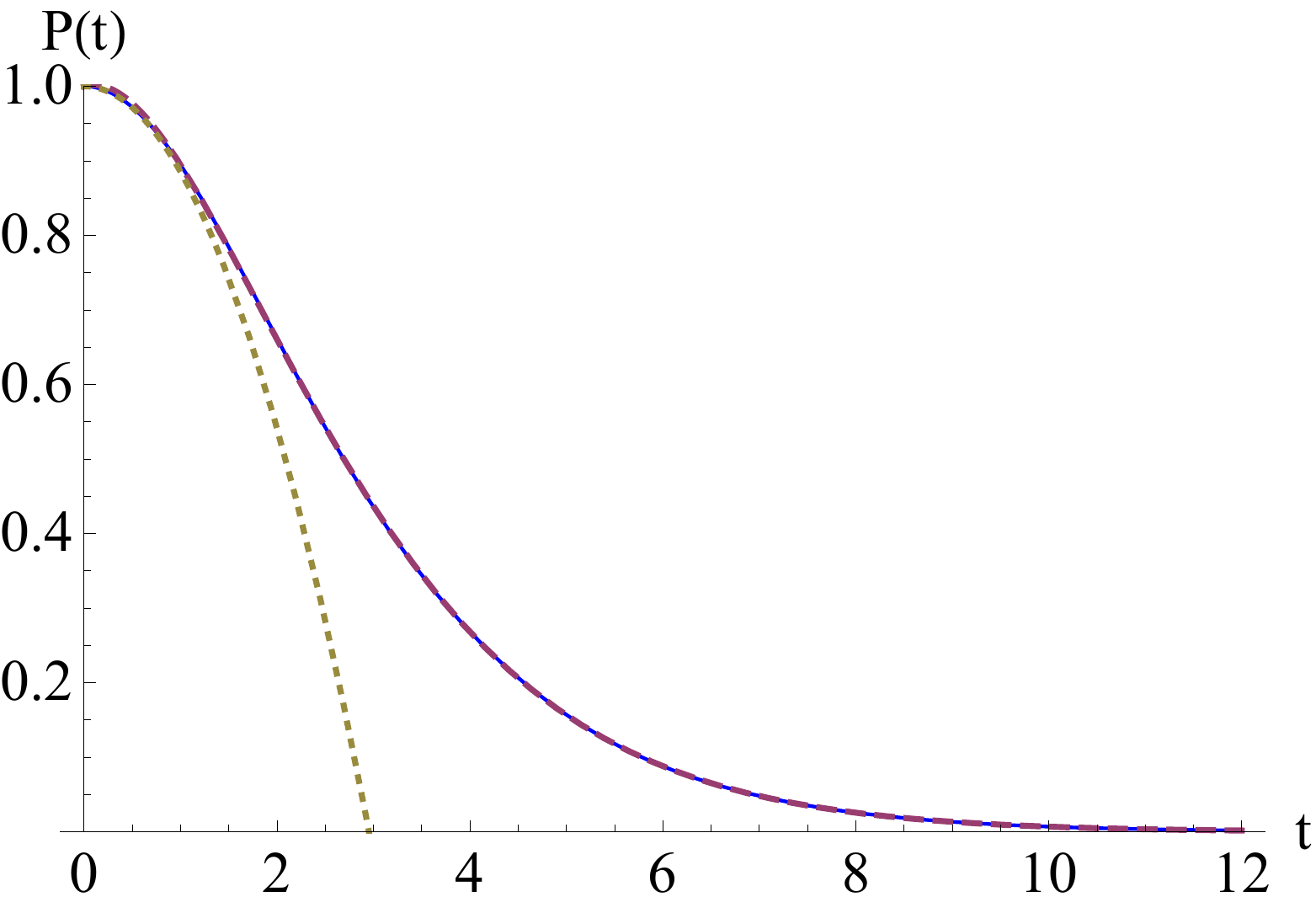}
\hfill
 \includegraphics[width=0.4\textwidth]{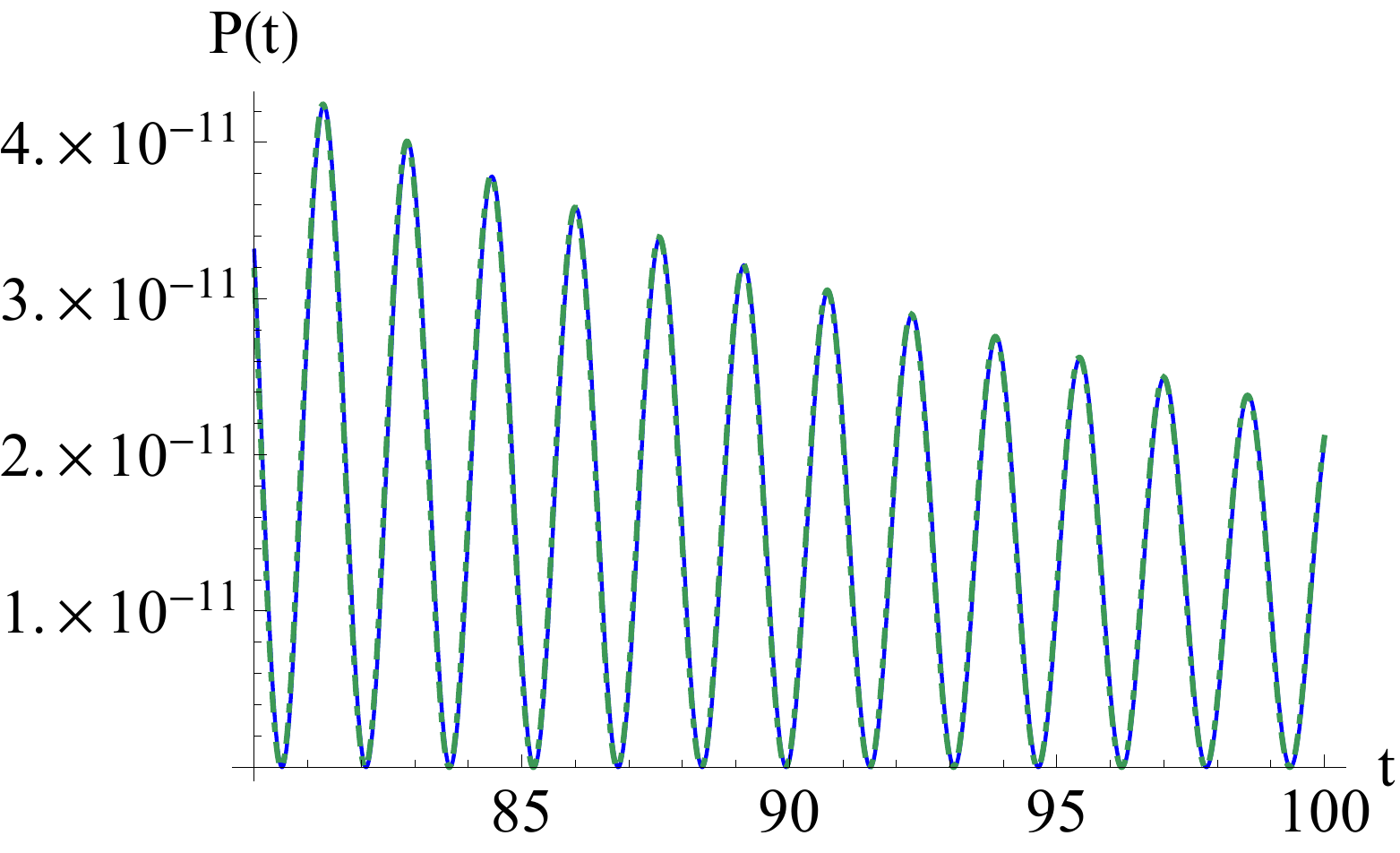}
 \hspace*{0.05\textwidth}
 \\
\hspace*{0.01\textwidth}(c)\hspace*{0.47\textwidth}(d)\hspace*{0.4\textwidth}
\\
\caption{(color online) (a) Linear plot  of the survival probability $P(t)$ near to the EP2B.  The solid blue curve gives the numerical integration of Eq. (\ref{II.GEP.surv}) while the purple dashed  curve represents the approximate expression $P_\textrm{P}(t)$ reported in Eq. (\ref{II.GEP.surv.EP}) for the intermediate timescale (the approximation is so accurate that the purple dashed curve might be difficult to distinguish from the blue curve).  Meanwhile, the beige dotted curve represents the short-time approximation from Eq. (\ref{II.surv.amp.zeno}).
Here $g=0.1$, such that the EP2B is located at $V=\VB= 0.00501256$ with $\eB=0.00502517 i$.  We have chosen  $V=0.00501260$, very close to $\VB$.
  (b) Linear plot of the survival probability for long times, exactly at the EP2B for $g=0.1$. The solid blue line is obtained from the numerical integration of Eq. (\ref{II.GEP.surv.EP.sum}) and the green chained line corresponds to the long-time approximation in Eq. (\ref{eq:EP2BLongT4}). 
  The plots (c) and (d) are the same as (a) and (b), respectively, but with different  parameters: here $g=0.75$, such that $\VB=0.3385620$  and $\eB=0.868647 i$. For (c) we chose $V=0.3385622$, which is again very close to $\VB$,
  while in (d) we chose $V = \VB$ exactly at the EP2B for $g=0.75$. 
  }
\label{fig:EP2BSurvProb} 
\end{figure}

Next we turn to the intermediate timescale, which of course is strongly influenced in the vicinity of the EP2B.  
We again rely on the formalism in Ref. \cite{HO14},
according to which the survival amplitude is
\begin{eqnarray}
  A(t) 
     &	= & \bra \dA | e^{-i H_\textrm{II} t} | \dA \ket
     						\nonumber  \\
     &	= & \frac{1}{2 \pi i} \sum_{j = \{ s, \pm \}} 
		\int_\mathcal{C} d \lambda \left( - \lambda + \frac{1}{\lambda} \right) 
			\exp \left[ i \left( \lambda + \frac{1}{\lambda} \right) t \right]
				\bra \dA | \psi_j \ket \frac{\lambda_j}{\lambda - \lambda_j} \bra \tilde{\psi}_j | \dA \ket
\label{II.GEP.surv}
\end{eqnarray}
where the sum over $j = \{ s, \pm \}$ incorporates all four sign combinations.
In the vicinity of the EP2B, we make use of the expansions for the eigenvalue and the eigenvector component from Eqs. (\ref{II.lambda.j.exp}) and (\ref{II.GEP.eigenket.dA.exp}) to write the sum appearing in Eq. (\ref{II.GEP.surv}) as
\begin{eqnarray}
  \sum_{j = \{ s, \pm \}} 
	\frac{\lambda_j}{\lambda - \lambda_j} \bra \dA | \psi_j \ket^2
	     &	=  & - \frac{ 1 + \left( 1-g^2 \right) \lambda^2 }
					{\left( 1 + \sqrt{1-g^2} \lambda^2 \right)^2} + O (\delta)
									\nonumber  \\
	     & = & \frac{\lamB^4 + \lambda^2}{\left( \lamB^2 - \lambda^2 \right)^2} + O (\delta)
		.
\label{II.GEP.surv.factor}
\end{eqnarray}
The survival amplitude then takes the form (to lowest order in $\delta$)
\begin{eqnarray}
  A(t) 
  	= \frac{1}{2\pi i} 
		\int_\mathcal{C} d \lambda \left( - \lambda + \frac{1}{\lambda} \right) 
			\exp \left[ i \left( \lambda + \frac{1}{\lambda} \right) t \right]
				\frac{\lamB^4 + \lambda^2}{\left( \lambda^2 - \lamB^2 \right)^2}
		.
\label{II.GEP.surv.EP.sum}
\end{eqnarray}
Deforming the contour and then taking the residue of the two terms for the EP $\lambda = +\lamB$ associated with the decaying solutions gives the pole contributions as
\begin{eqnarray}
  A_\textrm{P}(t) 
     & 	= & \lamB^4 \lim_{\lambda \rightarrow \lamB} \frac{d}{d\lambda} \left[
		\frac{- \lambda + \frac{1}{\lambda}}{\left( \lamB + \lambda \right)^2}
			e^{i \left( \lambda + \frac{1}{\lambda} \right) t} \right]
		+ \lim_{\lambda \rightarrow \lamB} \frac{d}{d\lambda} \left[
			\frac{\lambda \left( 1 - \lambda^2 \right)}{\left( \lamB + \lambda \right)^2}
				e^{i \left( \lambda + \frac{1}{\lambda} \right) t} \right]
							\nonumber  \\
     &	= & 
	 - \left[ 1 + \frac{t g^2 \left( 1 + \sqrt{1 - g^2} \right)}{4 \left( 1 - g^2 \right)^{3/4}} \right]
     		e^{- \Gamma t / 2}
\label{II.GEP.surv.EP.res}
\end{eqnarray}
Notice this result is purely real.  So then the survival probability at the EP2B on intermediate timescales is given by
\begin{eqnarray}
  P_\textrm{P}(t) 
     	= \left[ 1 + \frac{g^2 \left( 1 + \sqrt{1 - g^2} \right)}{2 \left( 1 - g^2 \right)^{3/4}} t
			+ \frac{g^4 \left( 2 -g^2 + 2 \sqrt{1 - g^2} \right)}{16 \left( 1 - g^2 \right)^{3/2}} t^2  \right]
     		e^{- 2 t \sqrt{\frac{2-g^2}{\sqrt{1 - g^2}} - 2}}
	.
\label{II.GEP.surv.EP}
\end{eqnarray}
This result gives a very good approximation for the survival probability on intermediate timescales over a wide range of $g$ values; a comparison with the numerical integration is shown for the case $g=0.1$ in Fig. \ref{fig:EP2BSurvProb}(a) and $g=0.75$ in \ref{fig:EP2BSurvProb}(c). We will discuss the reason for the extremely close agreement between the pole approximation in Eq. (\ref{II.GEP.surv.EP}) and the numerically-integrated survival probability at the end of next subsection.


\subsection{Inverse power law evolution on long timescales near the EP2B}\label{sec:II.long}

To calculate the survival probability for long times we will start with Eq. (\ref{II.GEP.surv.EP.sum}). Using the partial-fraction decomposition
\begin{eqnarray}
				\frac{\lamB^4 + \lambda^2}{\left( \lambda^2 - \lamB^2 \right)^2}
				=
				\frac{1}{4}\left[\left( \lamB^2+1\right)\left(\frac{1}{\left(\lam- \lamB\right)^2}
				+ \frac{1}{\left(\lam+ \lamB\right)^2}\right)
				+\left(\frac{1}{\lamB}-\lamB\right)\left(\frac{1}{\lam-\lamB}
				-\frac{1}{\lam+\lamB}\right)\right]
\label{eq:EP2BLongT1}
\end{eqnarray}
the survival amplitude takes the form
\begin{eqnarray}
				A(t) 
				&=&
				\frac{1}{4}\left[\left( \lamB^2+1\right)\frac{\partial}{\partial \lamB}  
				+\left(\frac{1}{\lamB}-\lamB\right)\right] 
				 \frac{1}{2\pi i} 
		\int_\mathcal{C} d \lambda \left( - \lambda + \frac{1}{\lambda} \right) 
			\exp \left[ i \left( \lambda + \frac{1}{\lambda} \right) t \right]
				\left(\frac{1}{\lam-\lamB}-\frac{1}{\lam+\lamB}\right)\nonumber\\
				&= &
				\frac{1}{4}\left[\left( \lamB^2+1\right)\frac{\partial}{\partial \lamB}  
				+\left(\frac{1}{\lamB}-\lamB\right)\right] \left( \II_\textrm{L} (\lamB,t) +  \II_\textrm{U} (\lamB,t) - \II_\textrm{L} (-\lamB,t) -  \II_\textrm{U} (-\lamB,t) \right)
			+ A_P(t)
\label{eq:EP2BLongT2}
\end{eqnarray}
where in the second line we 
follow a development similar to that appearing in Eqs. (\ref{I.surv.int.bp}) through (\ref{I.surv.int.bp.final}) for Model I and $A_P(t)$ is the contribution from the residue at the $\lambda=\lamB$ pole given in Eq. (\ref{II.GEP.surv.EP.res}).
Note that in contrast to the EP2A case in Model I, here we have to include contributions associated with both the lower and upper edges of the energy band, because the (pure imaginary) EP2B eigenvalue lies directly in the middle of the band on the real energy axis. 
Inserting the explicit expressions for all the long-time approximations to the integrals we have
\begin{eqnarray}
				A(t) 
				&\approx&
				\frac{1}{4}\left[\left( \lamB^2+1\right)\frac{\partial}{\partial \lamB}  
				+\left(\frac{1}{\lamB}-\lamB\right)\right] \nonumber\\
				&\times& \left[\frac{1}{t^{3/2}}\frac{1}{2\sqrt{\pi}}
				\left(
				    \frac{e^{2it +i \pi/4}}{-2+\eB} 
				+  \frac{e^{-2it - i\pi/4}}{2+\eB} 
				-  \frac{e^{2it +i \pi/4}}{-2-\eB}
				-  \frac{e^{-2it -i \pi/4}}{2-\eB}
				\right)  \right] + A_P(t)
\label{eq:EP2BLongT3}
\end{eqnarray}
where $\eB= \lamB + 1/\lamB$. The long-time approximations of $\II_\textrm{X} (\lamB,t)$ (with $\textrm{X}=\textrm{L}$ or $\textrm{R}$) in Eq. (\ref{eq:EP2BLongT3}) are valid when $t\gg T_{\rm EP,X}$, where
\begin{eqnarray}
				T_{\rm EP,L} = \frac{1}{|\eB +2|}, \qquad T_{\rm EP,R} = \frac{1}{|\eB -2|}
\label{eq:EP2BLongTEP}
\end{eqnarray}
For small $g$, these two times are both close to $1/2$ because the EP2B eigenvalue satisfies
\begin{eqnarray}
				\eB \approx i g^2/2,	
\label{eq:EP2BEB}
\end{eqnarray}
which can be seen by expanding Eq. (\ref{II.EP2B}).  Hence $T_{\rm EP,L} \approx T_{\rm EP,R} \approx 1/2$. This means that even for relatively short times (e.g. when the pole contribution is still large), Eq. (\ref{eq:EP2BLongT3}) is already applicable.

Simplifying Eq. (\ref{eq:EP2BLongT3})  we obtain
\begin{eqnarray}
				A(t) 
				&\approx&  \frac{1}{t^{3/2}}\frac{1}{2\sqrt{\pi}} \cos\left(2t + \pi/4\right)
				\left(\frac{1}{\lamB}-\lamB\right)
				  \frac{-2\eB^3}{4-\eB^2}	+ A_P(t)
\label{eq:EP2BLongT4}
\end{eqnarray}
After several exponential lifetimes $1/\gamB$
of the pole term $A_P(t)$
have been exhausted \cite{Sudarshan}, then the non-exponential $t^{-3/2}$ term gives the larger contribution to Eq. (\ref{eq:EP2BLongT4});
from this point forward the survival  probability decreases in time as $P(t) \sim t^{-3} \cos^2\left(2t + \pi/4\right) $, as shown in figures \ref{fig:EP2BSurvProb}(b) and (d).  Note that the oscillatory behavior in the survival probability is a result of interference between the contributions from the two band edges; similar oscillatory expressions are obtained for the long time dynamics in the middle of the band (but, unlike the present case, away from any exceptional points) in Refs. \cite{Longhi06,Zueco16}.

To make a more precise statement, we note that the $\eB^3$ factor in Eq. (\ref{eq:EP2BLongT4}) together with  Eq. (\ref{eq:EP2BEB}) for small $g$ implies that the $t^{-3/2}$ term of the survival amplitude  is proportional to $g^{6}$. This high-power dependence on $g$ implies that  when $t$ is {\it not} much larger than the relaxation time, the $t^{-3/2}$ term in  Eq. (\ref{eq:EP2BLongT4}) is still negligible in comparison with the pole term $A_P(t)$. This explains why the pole contribution gives such a close fit to the complete, numerically-integrated survival probability in Fig. \ref{fig:EP2BSurvProb} (a) with $g=0.1$. The fit is still very good even for $g=0.75$ as shown in Fig. \ref{fig:EP2BSurvProb} (c).


\section{Conclusion}\label{sec:conc}

We have studied the time evolution characteristics of open quantum systems in the vicinity of two coalescing eigenstates; while previous works have generally used a heuristic effective Hamiltonian method to study this behavior, here we have relied on an analytic approach starting from the microscopic Hamiltonian.  One strong advantage of our analysis is that while the heuristic method ignores the details of the structured environment in a given open system, we have been able to reveal that the continuum threshold can influence the dynamics in a variety of circumstances.

The better studied case in the literature is the EP2B, at which two resonance states coalesce before forming two different resonance states.  In this case, the eigenvalue at the exceptional point is still complex-valued, and hence the effective Hamiltonian method \cite{WKH08,CM11} predicts that the usual exponential decay process $e^{-\Gamma t}$ is still present in the survival probability, but modified with a term $t^2 e^{-\Gamma t}$ due to the double pole appearing in the survival amplitude.  This has been verified by experiment in Refs. \cite{EPexpt1d,TUD14}.  However, relying on the microscopic analysis, in this paper we have demonstrated that the influence of the continuum threshold should still result in a $t^{-3}$ evolution on the asymptotic timescale even at the EP2B. 

Meanwhile, our method reveals that the influence of the continuum can dramatically modify the dynamics in the vicinity of an EP2A, at which two virtual bound states coalesce before forming a resonance/anti-resonance pair.  In this case, the heuristic approach [equivalent to the pole approximation appearing in Eq. (\ref{surv.amp.Q}) in the present work, which yields a non-unitary evolution] fails completely for the model considered in Sec. \ref{sec:surv}.  Instead, we found that when the EP2A appears near the continuum threshold the intermediate timescale dynamics are dramatically modified, resulting in a non-exponential evolution of the form 
$P(t) \sim 1 - C_1 \sqrt{t} + D_1 t$ in which the characteristic $\sqrt{t}$ term has the strongest influence on the dynamics 
\footnote{It may be interesting to contrast the time evolution predicted here for an EP2A in a Hermitian open quantum system with that predicted in [E.-M. Graefe, H. J. Korsch, and A. E. Niederle, Phys. Rev. A {\bf 82}, 013629 (2010)] for a {\it closed} quantum system that is explicitly non-Hermitian.  In the first model appearing in that paper, the authors study the time evolution near an EP2A in a $\mathcal{PT}$-symmetric dimer consisting of two coupled levels, one with a steady energy gain and the other with energy loss (the gain and loss terms could be understood as approximations of true reservoirs with very narrow bandwidth).  Since their system is explicitly non-Hermitian, there is no restriction that the dynamics must satisfy unitarity; hence in that case, the equivalent of the pole approximation {\it does} adequately describe the dynamics and the authors find the system follows an evolution of the form $1 - C_2 t + D_2 t^2$ for all $t$.  No 
$\sqrt{t}$ term appears in that case because their model does not include any structured reservoir (i.e., no continuum threshold exists).}.  However, even when the EP2A is well-separated from the band edge, the pole approximation was still useless as the dynamics instead follows parabolic decay for an extended period during which most of the dissipation occurs. In either case, the typical $t^{-3}$ power law evolution asserts itself on the asymptotic timescale $t \gg 1/\delEP$, in which $\delEP$ is the energy gap between the EP2A eigenvalue and the continuum threshold.


From this point, it remains for future work to propose specific experiments to observe the phenomena studied in this paper, particularly the $\sqrt{t}$ evolution for the EP2A near the threshold.  However, given recent precision experiments performed in coupled waveguide arrays, particularly the semi-infinite array studied in Ref. \cite{BIC_WG}, the models studied in this work may not be so distant from what can be achieved in the laboratory.  Along similar lines, one could also consider a many-body extension of the present work \cite{many_body_expt,LDV12,delCampo11}.

Another natural extension of the present study would be to consider how the evolution might be further modified in the case that an EP2A appears directly at the continuum threshold.  
At a glance, we might consider this question for Model I in the limit 
$g \rightarrow 0$; according to Eq. (\ref{eps.EP}), the EP2A then approaches the band edge at $E = -2$.  However, for Model I this limit is of course just the trivial limit in which the coupling vanishes: Eqs. (\ref{I.surv.prob.zeno}), (\ref{I.P.band.edge}), and (\ref{I.surv.prob.fz}) show that the the decoupled impurity site simply remains occupied with $P(t) = 1$ for all $t$ in this case.
However, non-trivial cases for which the EP2A appears directly at the threshold can be found in Refs. \cite{TGKP16,Heiss_zero}.  In this unusual situation, the EP2A also coincides with a localization transition, such that a bound state and a virtual bound state merge before splitting into a resonance/anti-resonance pair \footnote{Note it has been shown in Ref. \cite{GGH15} that a slightly different process can occur in the case of an EP2A at threshold in $\PT$-symmetric open quantum systems.}.
The time evolution in this case will be the subject of future work.

\section*{Acknowledgements}

The authors would like to thank Naomichi Hatano for helpful discussions; S. G. particularly thanks him for guidance on the Jordan block calculation.  We would also like to thank Kazuki Kanki, Ken-ichi Noba, Tomio Petrosky, Satoshi Tanaka, and Dvira Segal for useful discussions and encouragement.  S. G. acknowledges support from the Japan Society for the Promotion of Science (Fellowship Grant No. PE12057), a Young Researcher's Grant from Osaka Prefecture University, as well as
the Program to Disseminate the Tenure Track System from the Ministry of Eduction, Culture, Sports, Science, and Technology in Japan. G. O. acknowledges the Institute of Industrial Science at the University of Tokyo, the Department of Physical Science at Osaka Prefecture University, the Holcomb Awards Committee and the LAS Dean's office at Butler University for support of this work.


\appendix


\section{Topological properties of the EP2A}\label{app:models.encircle}

In this appendix we study the topological properties in the vicinity of an EP2A by demonstrating how the eigenstates are modified as we evolve the system parameters so as to encircle the EP2A in Model I.   We emphasize that the following analysis does not represent a true dynamical evolution around the exceptional point, but rather a quasi-stationary parametric evolution similar to that demonstrated for an EP2B in the microwave cavity experiments reported in Refs. \cite{EPexpt1a,EPexpt1c}.  In those papers, the researchers have simulated a topological encirclement of the EP2B by performing separate measurements of the cavity modes at a set of independent pairs of parameter values that surround the exceptional point; this is opposed to a dynamical evolution, in which one would instead populate a specific mode, evolve the system parameters around the exceptional point, and then measure the mode population at the end of the cycle (clearly in this case one would have to evolve the system and perform the measurement more quickly than the lifetime of either of the modes involved in the transition).  This latter type of evolution has just recently been reported in two experiments for the case of the EP2B \cite{RotterNat,XuNat}.  

Here we will only attempt to demonstrate the quasi-stationary parametric (or topological) evolution for the EP2A, although at the end of the calculation we will briefly comment on what the analysis for the genuine dynamical evolution might involve.  
We note that the encirclement of the EP2A has received relatively little discussion in the literature for either type of evolution with the exception of the experiments in Refs. \cite{TUD11,TUD14}, which provide an example where the topological encirclement of an EP2A is simulated after an effective mapping from an EP2B.  We will comment on this experiment briefly after our calculation below.

Our development below for the topological evolution around the EP2A is similar in spirit to the theoretical analysis for the EP2B appearing in Ref. \cite{EPexpt1c}.  First we find it useful to rewrite the $\lambda$ eigenvalues from Eq. (\ref{I.lambda.pm}) as
\beq
  \lambda_\pm
  	= \xi_\textrm{d} \mp \sqrt{\xi_\textrm{d}^2 - \lamEP^2}
	= \xi_\textrm{d} \mp \sqrt{ \left( \xi_\textrm{d} + \lamEP \right)  \left( \xi_\textrm{d} - \lamEP \right)}
	.
\label{I.lambda.pm.xi}
\eeq
in which we have reparameterized $\epsd$ in terms of
\beq
  \xi_\textrm{d}
  	\equiv - \frac{\epsd}{2 \left( 1 - g^2 \right)}
	.
\label{xi.d}
\eeq
Next, in order to track how the phase of the eigenstates evolve as we encircle the EP we introduce the angle variable $\theta$ such that the $\lambda_-$ eigenvalue is written
\beq
  \frac{\lambda_- \left( \theta \right)}{\lamEP} = i \tan \theta
	.
\label{tan.theta}
\eeq
Plugging this into $| \Psi_- \ket = \beta_- \left[ 1, \lambda_- \right]$, and taking into account the form of the norm Eq. (\ref{I.norm}), enables us to write the corresponding eigenstate as
\beq
  | \Psi_- \ket
  	= \left[  \begin{array}{c}
		\cos \theta	\\
		i \lamEP \sin \theta
		\end{array} \right]
	.
\label{I.Psi.EP.m}
\eeq
Recalling the identity $\lambda_+ \lambda_- (1 - g^2) = 1$ from Eq. (\ref{I.lambda.pm.ident}) we can use Eq. (\ref{tan.theta}) to write the $\lambda_+$ eigenvalue in terms of the $\theta$ parameterization as
\beq
  \frac{\lambda_+ \left( \theta \right)}{\lamEP} = -i \cot \theta
  	= i \tan \left( \theta \pm \pi/2 \right)
\label{cot.theta}
\eeq
with the corresponding form of the eigenvector
\beq
  | \Psi_+ \ket
  	= \left[  \begin{array}{c}
		\sin \theta	\\
		- i \lamEP \cos \theta
		\end{array} \right]
	.
\label{I.Psi.EP.p}
\eeq
On comparing Eqs. (\ref{cot.theta}) and (\ref{tan.theta}), we see that the $\lambda_\pm$ eigenvalues are related to each other through a phase shift of $\pm \pi/2$ as 
$\lambda_+ \left( \theta \right) = \lambda_- \left( \theta \pm \pi/2 \right)$.

Now let us deform the parameter $\xi_\textrm{d}$ to perform an encirclement of the lower EP2A $\epsEP$.  We could accomplish this most simply by varying $\xi_\textrm{d}$ on the trajectory $\xi_\textrm{d} = \lamEP + \chi e^{i \delta}$ as $\delta$ is taken from $0$ to $2 \pi$ (clockwise encirclement) or $0$ to 
$- 2 \pi$ (counter-clockwise).  From Eq. (\ref{I.lambda.pm.xi}) we see that the $\lambda_\pm$  eigenvalues will then encircle $\lamEP$ as long as $\chi/\lamEP < 1$ is satisfied (if $\chi/\lamEP > 1$ then both EPs are encircled and the branch cut has no effect).

Now let us further specify the parameter $\theta$ such that the counter-clockwise encirclement is given by 
$\theta \rightarrow \theta + \pi / 2$ while the clockwise case is given by $\theta \rightarrow \theta - \pi /2$.  We can observe how the encirclement of the exceptional point modifies the eigenvectors in either direction as
\begin{equation}
  | \Psi_- (\theta) \ket 
	\rightarrow
  | \Psi_- (\theta \pm \pi/2) \ket 
			= \left[ 
			\begin{array}{c}
		\cos \left( \theta	\pm \pi/2 \right)	 \\
		i \lamEP \sin \left( \theta \pm \pi/2 \right)
			\end{array} 
			\right]
		=
			\left[ 
			\begin{array}{c}
		\mp \sin \theta				 \\
		\pm i \lamEP \cos \theta
			\end{array} 
			\right]
		=
		\mp | \Psi_+ (\theta) \ket
\label{I.Psi.EP.m.encircle}
\end{equation}
and
\begin{equation}
  | \Psi_+ (\theta) \ket 
	\rightarrow
  | \Psi_+ (\theta \pm \pi/2) \ket 
			= \left[ 
			\begin{array}{c}
		\cos \left( \theta	\pm \pi/2 \right)	 \\
		- i \lamEP \sin \left( \theta \pm \pi/2 \right)
			\end{array} 
			\right]
		=
			\left[ 
			\begin{array}{c}
		\pm \sin \theta				 \\
		\pm i \lamEP \cos \theta
			\end{array} 
			\right]
		=
		\pm | \Psi_- (\theta) \ket
		.
\label{I.Psi.EP.p.encircle}
\end{equation}
Hence in the counterclockwise evolution the eigenvectors are rotated into one another, but 
$| \psi_- \ket$ picks up a minus sign as $| \psi_- \ket \rightarrow - | \psi_+ \ket $.  Meanwhile for the clockwise case we have the same result except that it is $| \psi_+ \ket$ that picks up the minus sign as $| \psi_- \ket \rightarrow | \psi_+ \ket$ and $| \psi_+ \ket \rightarrow - | \psi_- \ket$.  As a result, the system requires four revolutions in order for both the eigenvectors and the eigenvalues to return to the original configuration.  

The sequence of transformations for the counterclockwise rotation is therefore given as
\begin{equation}
			\left \{
			\begin{array}{c}
		| \Psi_+ \ket 	 \\
		| \Psi_- \ket 
			\end{array} 
			\right \}
		\circlearrowleft
			\left \{
			\begin{array}{c}
		| \Psi_- \ket 	 \\
		- | \Psi_+ \ket 
			\end{array} 
			\right \}
		\circlearrowleft
			\left \{
			\begin{array}{c}
		- | \Psi_+ \ket 	 \\
		- | \Psi_- \ket 
			\end{array} 
			\right \}
		\circlearrowleft
			\left \{
			\begin{array}{c}
		- | \Psi_- \ket 	 \\
		| \Psi_+ \ket 
			\end{array} 
			\right \}
		\circlearrowleft
			\left \{
			\begin{array}{c}
		| \Psi_+ \ket 	 \\
		| \Psi_- \ket 
			\end{array} 
			\right \}
\label{semi.quad.psi.rotation.ccw}
\end{equation}
while that for the clockwise case is given as
\begin{equation}
			\left \{
			\begin{array}{c}
		| \Psi_+ \ket 	 \\
		| \Psi_- \ket 
			\end{array} 
			\right \}
		\circlearrowright
			\left \{
			\begin{array}{c}
		- | \Psi_- \ket 	 \\
		| \Psi_+ \ket 
			\end{array} 
			\right \}
		\circlearrowright
			\left \{
			\begin{array}{c}
		- | \Psi_+ \ket 	 \\
		- | \Psi_- \ket 
			\end{array} 
			\right \}
		\circlearrowright
			\left \{
			\begin{array}{c}
		| \Psi_- \ket 	 \\
		- | \Psi_+ \ket 
			\end{array} 
			\right \}
		\circlearrowright
			\left \{
			\begin{array}{c}
		| \Psi_+ \ket 	 \\
		| \Psi_- \ket 
			\end{array} 
			\right \}
\label{semi.quad.psi.rotation.cw}
\end{equation}
This rotation property captured in Eqs. (\ref{semi.quad.psi.rotation.ccw}) and (\ref{semi.quad.psi.rotation.cw})
is a general mathematical property of EP2s (EP2A or EP2B) 
as confirmed in the previously mentioned quasi-stationary experiments for the EP2B \cite{EPexpt1a,EPexpt1c}, as well as for an {\it effective} EP2A in Refs. \cite{TUD11,TUD14}; however, it should be emphasized that in the latter case the EP2A is only obtained after the authors perform a shift on their Hamiltonian such that an EP2B is mathematically mapped into an EP2A (and hence most of the following discussion would not apply for that case).

We again point out that in the above calculation we have only studied how the mode structure evolves as we modify the system parameters; whereas for a genuine dynamical evolution we would instead need to consider how a given state vector evolves according to the Schr\"odinger equation for a specific time-dependent Hamiltonian.  This much more involved calculation is carried out for the EP2B in Refs. \cite{GMM13,Rotter15}, for example, in which it is demonstrated that the quasi-stationary method is insufficient to describe the dynamical evolution.  However, we note that in Refs. \cite{GMM13,Rotter15}, the authors again rely on a heuristic 2x2 effective Hamiltonian.  As we demonstrate in the main text of the present paper, that approach should be sufficient for the case of the EP2B (or the effectively mapped EP2A as in Refs. \cite{TUD11,TUD14}), as one should naturally choose an evolution faster than the lifetime of the states involved in any case (and as we demonstrate in Sec. \ref{sec:II.ham}, the threshold should usually have negligible influence on this timescale for the EP2B).
On the other hand, to study the dynamical evolution around a true EP2A in a Hermitian open quantum system, one would likely have to abandon the heuristic effective Hamiltonian and instead rely on a method that fully incorporates the influence of the continuum threshold on the evolution.  One would likely also have to consider that, generally speaking, the EP2A is more stable against perturbations of the system parameters than the EP2B, as pointed out in Ref. \cite{GGH15}.
While we have posed the question here, we will not consider this highly non-trivial issue further in the present work.



\section{Proof of Eq. (\ref{eq:xrep})}
\label{app:ProofSurvAmp}

The generalization of Eq. (\ref{eq:xrep}) has been proven for general tight-binding models in Ref. \cite{HO14}. Here we will sketch this derivation for Model I, presented in section \ref{sec:I} B.

Starting with Eq. (\ref{I.A.B}) we have
\beq
  	F-\lambda G = \left[ \begin{array}{cc}
	         -\lambda	&	1	\\
		1		&	H_{\rm eff}(E) +\lambda
	\end{array} \right]
		,
\label{eq:FmlG}
\eeq
and
\beq
  	(F-\lambda G)^{-1} = \frac{1}{\lambda (E-H_{\rm eff}(E))} 
	\left[ \begin{array}{cc}
	         H_{\rm eff}(E)+\lambda	&	-1	\\
		-1	&	-\lambda
	\end{array} \right]
		.
\label{eq:FmlGinv}
\eeq
Isolating the element on the first row and second column we have
\beq
  	(E-H_{\rm eff}(E))^{-1} 
	= -\lambda
	\left[ \begin{array}{cc}
	        1	&	0	
	     \end{array} \right]
	(F-\lambda G)^{-1} 
		\left[ \begin{array}{cc}
	         0\\
		1	
	\end{array} \right]
		.
\label{eq:GfFG}
\eeq
Next we turn to Eqs. (\ref{B.ident}) and (\ref{A.Lambda}), which give 
\beq
(F-\lambda G)^{-1}= U(\Lambda-\lambda I_2)^{-1}{\tilde U}.
\label{eq:FGLam}
\eeq
Furthermore we have
\beq
  	U = 
	\left[ \begin{array}{cc}
	         \bra d|\psi_+\ket	&	\bra d|\psi_-\ket	\\
		\lambda_+ \bra d|\psi_+\ket	&	\lambda_- \bra d|\psi_-\ket
	\end{array} \right]
\label{eq:Upsi}
\eeq
and
\beq
  	{\tilde U} = 
	\left[ \begin{array}{cc}
	         \bra {\tilde \psi}_+|d\ket	&	\lambda_+  \bra {\tilde \psi}_+|d\ket	\\
		\bra {\tilde \psi}_-|d\ket	&	\lambda_-  \bra {\tilde \psi}_-|d\ket
	\end{array} \right]
	.
\label{eq:Utpsi}
\eeq
 Eqs. (\ref{eq:GfFG})-(\ref{eq:Utpsi}) together with Eq. (\ref{eq:AHeff}) give Eq. (\ref{eq:xrep}), after changing the integration variable from $E$ to $\lambda$.

\section{Transformation of the integral $\II(\lam_n,t)$ and derivation of Eq. (\ref{eq:fn2})}
\label{appA}

To derive Eq. (\ref{eq:fn2}), we first start with the integral appearing in Eq. (\ref{eq:fn}) and change integration variables from $\lambda$ to $E = -(\lambda+1/\lambda)$, so that
\begin{align} 
&& \lambda = -\frac{E}{2} + i \sqrt{1-E^2/4} \nonumber\\
&& \frac{1}{\lambda} = -\frac{E}{2} - i \sqrt{1-E^2/4} 
\end{align}
Then we have
 \begin{align} \label{eq:Aj}
 \II(\lam_n,t) =\frac{1}{2\pi i}  \int_{C_E}dE\,e^{-i E t}
\frac{1/\lam_n}{ E/2 + i \sqrt{1-E^2/4} +1/\lambda_n}
\end{align}
where $C_E$ is a counter-clockwise contour on the complex $E$ plane, surrounding the branch cut along $E \in [-2,2]$; see Fig. \ref{fig:contCE}(a).  Note that, more generally, we would also have to include a pole for any bound states present in the first Riemann sheet; however, for our present choice of the system parameters, we can assume there are none.  Even in the presence of bound states, the qualitative features below and in Sec. \ref{sec:surv.int2} for the influence of the EP2A on the dynamics should be similar (though possibly less obvious features in the survival probability in that case).

Multiplying and dividing the integrand in Eq. (\ref{eq:Aj})  by $E/2+1/\lambda_n-i\sqrt{1-E^2/4}$ we obtain
 \begin{align} 
\II(\lam_n,t)  =\frac{1}{2\pi i}
\int_{C_E}dE\,e^{-i E t} 
\frac{E/2 +1/\lambda_n - i \sqrt{1-E^2/4}}{ E-E_n}
	.
\end{align}
The integration around the branch-cut is given by
 \begin{align} 
\II(\lam_n,t)   
& = - \frac{1}{2\pi}
\int_{C_E} dE\,e^{-i E t} 
\frac{\sqrt{1-E^2/4}}{ E-E_n}		\nonumber   \\
& = - \frac{1}{2\pi}
\left( \int_{C+} + \int_{C-}\right) dE\,e^{-i E t} 
\frac{\sqrt{1-E^2/4}}{ E-E_n}
	\label{I.int.E.app}
\end{align}
where $C+$ is the contour just above the branch cut, going from $E=2$ to $E=-2$ and $C-$ is the contour just below the branch cut, going from $E=-2$ to $E=2$. The square root term changes sign in going from just above to just below the cut. Therefore we have
 \begin{align} 
\II(\lam_n,t) 
=
\frac{1}{\pi} \int_{-2}^2 dE\,e^{-i E t} 
\frac{\sqrt{1-E^2/4}}{ E-E_n}
\end{align}

Hereafter we will assume that  $E_n$  has a negative imaginary part. If the imaginary part is actually zero (for a virtual bound state), we can add an infinitesimal imaginary part $-i \epsilon$ to $E_n$ and then take the limit $\epsilon \to 0$ at the end. If the imaginary part of $E_n$ is positive (for an anti-resonance state), then the integration over $\tau$ in Eq. (\ref{Itau}) below should be taken from $0$ to $-\infty$. Assuming ${\rm Im}(E_n)<0$ is satisfied we write $\II(\lam_n,t) $ as \cite{OH}
 \begin{align} \label{Itau}
\II(\lam_n,t)  =-\frac{i }{\pi} \int_0^\infty d\tau\int_{-2}^2 dE\,e^{-i E t} 
e^{i(E-E_n)\tau}\sqrt{1-E^2/4} 
\end{align}
With the change of variable $E\to k$ such that $E=-2\cos(k)$ we have
 \begin{align} 
\II(\lam_n,t)  =-\frac{i }{\pi} \int_0^\infty d\tau e^{-iE_n\tau} \int_0^\pi  dk\,  2 \sin^2(k)e^{2i(t-\tau)\cos(k)}
 \end{align}
The integral over $k$ can be written in terms of Bessel's function $J_1$:
 \begin{align} 
 \II(\lam_n,t) =  - i  \int_0^\infty d\tau e^{-iE_n\tau} \frac{J_1[2(t-\tau)]}{t-\tau}
 \end{align}
Now we  change the integration variable $\tau$ to $t'=t-\tau$: 
 \begin{align} 
\II(\lam_n,t) &=  - i  \int_{-\infty}^t dt' e^{-iE_n(t-t')} \frac{J_1[2t']}{t'} \nonumber\\
 &=  -i  e^{-iE_n t} \left(\int_{-\infty}^0 dt' +  \int_{0}^{t} dt' \right) e^{iE_n t'} \frac{J_1[2t']}{t'}
\end{align}
The integral from $-\infty$ to $0$ can be evaluated exactly in terms of the hypergeometric function ${}_2F_1$:
 \begin{align} 
\label{HyperGF}
&\int_{-\infty}^0  dt' e^{iE_n t'} \frac{J_1[2t']}{t'}  
=\frac{1}{ i E_n}\frac{\Gamma(1)}{\Gamma(2)} {}_2F_1  \left[\frac{1}{2},1; 2; \frac{4}{E_n^2}\right] 
=\frac{1}{ i E_n} \left[\frac{1}{2} + \frac{1}{2} \sqrt{1-\frac{4}{E_n^2}}\right]^{-1}
= i \lambda_n^s
\end{align}
where $s=1$ if  $|\lambda_n|<1$, and $s=-1$ otherwise. Therefore we obtain
 \begin{align} \label{eq:I9}
& \II(\lam_n,t)  =  e^{-iE_nt}\left[ \lambda_n^s 
 -i  \int_{0} ^t dt' \, e^{i E_n t'}\frac{J_1(2t')}{t'} \right]
\end{align}
which gives Eq. (\ref{eq:fn2}) in the main text with $s=-1$ and $\lam_n=\lam_0$.

\section{Integral of Bessel's function} 
\label{appB}

When $E_n$ is near a band edge (say the edge at $E=-2$) we can  make an expansion of the integral in Eq. (\ref{eq:I9}) in terms of the variable $\Delta_n=-(E_n+2)$ \cite{GPSS13}:
 \begin{align} \label{eq:I10}
& \int_{0} ^t dt' \, e^{i E_n t'}\frac{J_1(2t')}{t'} = \int_{0} ^t dt' \, e^{-2i t'}\frac{J_1(2t')}{t'} e^{-i t'\Delta_n} 
=\sum_{l=0}^\infty \frac{(-i \Delta_n)^l}{l!} K_l(t)
\end{align}
where
 \begin{align} \label{eq:I11}
K_l(t) =  \int_{0} ^t dt' \, e^{-2i t'}\frac{J_1(2t')}{t'} (t')^l
	.
\end{align}
This gives
 \begin{align} \label{eq:I12}
& \II(\lam_n,t)  = \lam_n e^{-iE_nt}\left[ \lambda_n^s 
 -i \sum_{l=0}^\infty \frac{(-i \Delta_n)^l}{l!} K_l(t) \right]
 	.
\end{align}
The integrals $K_l(t)$ can be evaluated recursively. The first few are
 \begin{align} \label{eq:K0}
K_0(t) =  i\left[-1 +  e^{-2it}  \left(J_0(2t) + i J_1(2t)\right) \right]
	,
\end{align}
 \begin{align} \label{eq:K1}
K_1(t) =  \frac{1}{2}\left[1 - e^{-2it}  \left((1+2it) J_0(2t) -2t J_1(2t)\right) \right]
	,
\end{align}
 \begin{align} \label{eq:K2}
K_2(t) =  \frac{t}{3}e^{-2it}\left[ -it J_0(2t) +(i+t)  J_1(2t)\right]
	,
\end{align}
 \begin{align} \label{eq:K22}
K_3(t) =  \frac{t}{10}e^{-2it}\left[ -t(1+2it) J_0(2t) +  (1+2it+t^2) J_1(2t)\right]
	.
\end{align}
Note that in Eq. (\ref{eq:I12}), for each $l$, the highest power in $t$ outside of the Bessel functions appears as $(\Delta_n t)^l$.

If we assume that $|\Delta_n|\ll1$, meaning that the eigenvalue $E_n$ is close to the branch point at $E=-2$, 
we can keep just the zeroth-order term $K_0(t)$: 
 \begin{align} \label{eq:I13A}
\II(\lam_n,t)  &\approx \lam_n e^{-iE_nt}\left[ \lambda_n^s 
 -i  K_0(t) \right]\nonumber\\
 &= \lam_n e^{-iE_nt}\left[ \lambda_n^s -1\right] + \lam_n e^{-i(E_n+2)t} \left(J_0(2t) + i J_1(2t)\right) 
 	.
 \end{align}
This result together with Eq. (\ref{eq:BAsymp}) confirms the result \cite{GPSS13} that the  survival amplitude has a $1/\sqrt{t}$ dependence for the time scale $1\ll t \ll 1/|\Delta_n|$, as long as $E_n$ is not close to the EP. However, when $E_n$ is close to the EP, the result is modified as in Eq. (\ref{eq:BAsymp2}). 


\end{document}